\documentclass[12pt]{article}
\newmuskip\pFqmuskip
\newcommand*\pFq[6][8]{%
	\begingroup 
	\pFqmuskip=#1mu\relax
	\mathcode`\,=\string"8000
	\begingroup\lccode`\~=`\,
	\lowercase{\endgroup\let~}\pFqcomma
	{}_{#2}F_{#3}{\left(\left.\genfrac..{0pt}{}{#4}{#5}\right|#6\right)}%
	\endgroup
}
\newcommand{\pFqcomma}{\mskip\pFqmuskip}
\newcommand{\circo}{~\raisebox{1pt}{\tikz \draw[line width=0.6pt] circle(1.1pt);}~}
\pdfoutput=1
\usepackage{graphicx}
\usepackage{jheppub}
\usepackage{tikz}
\usepackage{hyperref}
\usepackage{amsmath}

\newenvironment{psmallmatrix}
{\left[\begin{smallmatrix}}
	{\end{smallmatrix}\right]}

\title{RG flow  between  $W_3$ minimal models by  perturbation and   domain wall approaches}

\usepackage{mathtools}

\DeclarePairedDelimiterX\braket[2]{\langle}{\rangle}{#1 \delimsize\vert #2}

\author[]{Hasmik Poghosyan }
\author[]{Rubik Poghossian}

\affiliation[]{Yerevan Physics Institute\\
Alikhanian Br. 2, 0036 Yerevan, Armenia}

\emailAdd{h.poghosyan@yerphi.am}
\emailAdd{poghos@yerphi.am}

\abstract{
We explore the RG flow 	between neighboring minimal  CFT models  with $W_3$ symmetry.
After computing several  classes of OPE structure
constants we were able to find the   matrices of anomalous dimensions for three classes of RG invariant  sets of local 
fields.
Each set from the first class consists of a single primary field, the second  one  of  three primaries, 
while  sets in the third  class contain  six primary and four secondary fields. We  diagonalize  their matrices of  anomalous dimensions 
and   establish the explicit  maps between UV and IR fields  (mixing coefficients).

While investigating the three point functions of secondary fields we have encountered an 
interesting phenomenon, namely  violation of  holomorphic anti-holomorphic factorization property,
something that does not happen in   ordinary minimal models with  Virasoro symmetry solely.

Furthermore,  the perturbation under consideration preserves a non-trivial subgroup of  $W$ transformations.
We have derived the corresponding conserved current   explicitly. 
We used this current to define  a notion of anomalous $W$-weights in perturbed theory: the analog for  matrix of 
anomalous dimensions. For RG invariant sets with primary fields only we have derived a formula for this quantity in terms of structure constants. This
allowed us to compute  anomalous $W$-weights for the first and second classes explicitly.

The same RG flow we investigate also with the domain wall approach for the second RG invariant class and find complete agreement with the 
perturbative approach. 
}

\begin{document}
\maketitle
\newcommand{\ie}{{\it i.e.\ }}
\def\bea{\begin{eqnarray}}
\def\eea{\end{eqnarray}}

\section*{Introduction}
In \cite{Zamolodchikov:1987ti}  A.~Zamolodchikov
investigated the RG flow from minimal model
${\cal M}_p$  to  ${\cal M}_{p-1}$ initiated by the
relevant field $\phi_{1,3}$. Using leading order perturbation
theory valid for $p\gg1$, for  several classes of local fields
he calculated the mixing coefficients specifying
the UV - IR map. The next to leading order
perturbation was analyzed in \cite{Poghossian:2013fda}. 
A similar 
RG trajectory connecting ${\cal N}=1$ super-minimal models
${\cal SM}_p$  to  ${\cal SM}_{p-2}$ was found in \cite{Poghossian:1987ngr} (see also \cite{Kastor:1988ef, Crnkovic:1989gy,Ahn:2014rua}). 
In this case the RG flow is initiated by the top component of the Neveu–Schwarz 
super-field $\Phi_{1,3}$.

For us it will be important that such a RG trajectory exists \cite{lukyanov1991additional}
also in the case of minimal models $A_{n-1}^{(p)}$ exhibiting $W_n$ extended conformal symmetry \cite{Zamolodchikov:1985wn}.
Namely, under certain slightly relevant perturbation  the theory  $A_{n-1}^{(p)}$ 
flows to $A_{n-1}^{(p-1)}$. 
In contrary to previous cases, this RG trajectory is not much investigated and 
we hope that in case of $n=3$ the present work will substantially fill this gap. 
After computing several  classes of OPE structure
constants we construct the matrices of anomalous dimensions for three RG invariant classes of local 
fields and establish the detailed pattern of UV/IR map in these sectors.   

To find the matrices of anomalous dimensions we need  several  classes of OPE structure constants.
Some of them were already derived  for Toda CFTs in \cite{Fateev:2007ab}. 
Even for these known cases the analytic   continuation to the minimal models of our interest is quite subtle.
This is why we have preferred to derive these and some previously unknown structure constants from scratch. 
We consider three RG invariant classes and derive corresponding matrices of anomalous dimensions. 
The first class consists of  a single primary, the second one  three primaries and the third   includes 
six primary and four level one secondary fields. 
In all cases we have  diagonalized  the matrices of  anomalous dimensions 
and identified those specific combinations of UV fields which flow to certain primaries of IR theory.

While investigating the three point functions of secondary fields we have encountered an 
interesting phenomenon, namely  violation of  holomorphic anti-holomorphic factorization property,
something that never happens in ordinary Virasoro minimal models, see (\ref{a_and_aa}). This does not 
contradict general principles since $W$ algebra symmetry generically speaking is not strong enough to reduce 
correlators of secondary fields to those with primaries only. 

We have shown that the perturbation under consideration preserves a 
subgroup of $W$ transformations and constructed the corresponding conserved 
current explicitly. Based in this analysis we introduce  and investigate the notion of 
anomalous $W$-weights (\ref{Def_WanomDim}) in close analogy with anomalous dimensions. 
Using our definition  we derive an  elegant expression 
for this matrix in terms of OPE structure constants (\ref{QmatEl}). This formula
holds  when the fields are primary which is not the case for  third class.

In many cases the UV/IR mixing coefficients can be computed using a completely different
method  by constructing the corresponding  RG domain wall.
Thus in \cite{Fredenhagen:2005an, Brunner:2007ur} it was suggested that there exists an interface 
that encodes the map from UV observables to IR.
In \cite{Brunner:2007ur} the RG domain wall was constructed for the $N=2$ super-conformal
models using matrix  factorization technique.

A nice algebraic construction of RG domain walls for coset CFT models was proposed in \cite{Gaiotto:2012np}. 
For the  case of Virasoro minimal models it was shown that this domain wall correctly reproduces 
perturbative results of \cite{Zamolodchikov:1987ti}. Later the consequences of this proposal 
has been carefully tested in various situations (see e.g.
\cite{Poghosyan:2013qta,Poghossian:2013fda, Poghosyan:2014jia, Konechny:2014opa}).

The seminal example of RG flow in \cite{Zamolodchikov:1987ti} has been extended for a very large 
class of CFT coset models
\begin{eqnarray}
	\mathcal{T}_{UV}&=&\frac{\hat{g}_l \times \hat{g}_m}{\hat{g}_{l+m}}\,, \qquad m>l
	\label{GenUV}
\end{eqnarray}
in  \cite{Crnkovic:1989gy,Ravanini:1992fs}, where it was argued that
under perturbation by a relevant field  (denoted by $\phi=\phi^{Adj}_{1,1}$),  
$\mathcal{T}_{UV}$ flows to the theory
\begin{eqnarray}
	\mathcal{T}_{IR}&=&\frac{\hat{g}_l \times \hat{g}_{m-l}}{\hat{g}_{m}} \,.
	\label{GenIR}
\end{eqnarray}
Since   $W_3$ minimal models $A_{2}^{(p)}$ belong to this class:
\bea 
\label{cosetrep}
A_{2}^{(p)}=\frac{\widehat{su}(3)_{p-3} \times \widehat{su}(3)_1}{\widehat{su}(3)_{p-2}}\,.
\eea 
It is natural to carry out an alternative analysis of RG 
trajectory $A_{2}^{(p)}\to A_{2}^{(p-1)}$ by explicit construction of Gaiotto's domain wall. 
As in \cite{Poghosyan:2014jia}, the method we applied is based on the current algebra construction
directly and, in this sense, is more general than the one originally employed in  \cite{Gaiotto:2012np}.
We have developed an appropriate higher rank generalization of this technique and applied it for the first 
non-trivial RG invariant set of three primaries, demonstrating its consistency with 
our perturbative result. Unfortunately calculations are rather cumbersome, and we have left the analysis 
of cases including descendants for a future work. Such investigation is very desirable, since 
the leading order perturbative analysis does not fix the mixing matrix completely due to 
inherent degeneracy in conformal dimensions. Another route to attack this problem would be 
extension of the notion of anomalous W-weights in case of descendants.

The paper is organized as follows: Section \ref{review} (besides  expressions for the 
structure constants (\ref{Cdiag}), (\ref{offdiag1})-(\ref{str_simple3})) is a review on CFTs with $W_3$ symmetry and the RG flow.
In section \ref{Matrix_of_anomalous} matrices of anomalous dimensions are derived for all three classes of RG invariant sets. 
In section \ref{Wanom} we introduce the notion of anomalous $W$-weights and derive its leading order   expression in terms of structure constants. This matrix is  constructed explicitly for the first two classes.  The RG domain wall is reviewed and explicitly constructed for the second class in section \ref{RGdomaiWall}. Finally many detailed computation can be found in the extensive appendices in this paper. In particular appendix \ref{SC} contains many important results which are instrumental in deriving the structure constants. In this appendix  we meet an intriguing absence  of holomorphic anti-holomorphic factorization property in three point functions including descendant fields.

\section{A review of $W_3$ minimal models and  RG flow}
\label{review}
\subsection{$W_3$  conformal field theories}
In any conformal field theory the energy-momentum tensor
has two nonzero components: the holomorphic field $T(z)$ with
conformal dimension $(2,0)$ and its anti-holomorphic
counterpart $\bar{T}(\bar{z})$ with dimensions $(0,2)$.
In conformal theories with extended  $W_3$ algebra summery one has in addition
the  currents $W(z)$ and $\bar{W}(\bar{z})$ with
dimensions $(3,0)$ and $(0,3)$ respectively.
These fields satisfy the OPE rules\footnote{The corresponding expressions for the anti-holomorphic counterparts  look exactly the same: one 
	simply substitutes $z$ by $\bar{z}$. This is why  we'll mainly concentrate on the
	holomorphic part.}
{\footnotesize
	\begin{eqnarray}
	\label{TTOPE}
	&&	T(z)T(0)=\frac{c/2}{z^4}+\frac{2T(0)}{z^2}+\frac{T^\prime(0)}{z}  + \cdots , 
	\\ \label{TGOPE}
	&&     T(z)W(0)=\frac{3W(0)}{z^{2}}+\frac{W^\prime(0)}{z} +\cdots\,,\\
	\label{GGOPE}
	&&	W(z)W(0)= \frac{c/3}{z^6}+\frac{2T(0)}{z^4} + \frac{T^\prime(0)}{z^3}+\frac{1}{z^2}\left(\frac{15c+66}{10(22+5c)}T^{\prime\prime}(0)+\frac{32}{22+5c}\Lambda(0)\right)\\
	&& \qquad \qquad\qquad\qquad \qquad\qquad\qquad \qquad
	+\frac{1}{z}\left(\frac{1}{15}T^{\prime\prime\prime}(0)+\frac{16}{22+5c}\Lambda^{\prime}(0)\right) +\cdots  .\nonumber
	\end{eqnarray}
}
Here the field $\Lambda(z)$ is  defined as
\bea
\label{Lambda}
\Lambda(z)=:\!\!TT\!\!:(z)-\frac{3}{10}T^{\prime\prime}(z)\,,
\eea
where $::$ is  regularization by means of subtraction of all  OPE singular terms. 
Without the second term in (\ref{Lambda}) 
 $\Lambda(z)$ wouldn't  be   quasi-primary. The state created by this field 
\bea
\Lambda(0)|0\rangle=\left(L_{-2}^2-\frac{3}{10}L_{-1}^2L_{-2}\right)|0\rangle
\eea 
indeed is a  Virasoro quasi-primary state i.e. $L_1\Lambda(0)|0\rangle=0$.
We can expand these fields in Laurent series
\begin{equation}
T(z)=\sum \limits_{n=-\infty}^{+\infty} \frac{L_n}{z^{n+2}}\,,
\quad
W(z)=\sum \limits_{n=-\infty}^{+\infty} \frac{W_n}{z^{n+3}}\,,
\quad
\Lambda(z)=\sum \limits_{n=-\infty}^{+\infty} \frac{\Lambda_n}{z^{n+4}}\,.
\end{equation}
The OPEs  (\ref{TTOPE}), (\ref{TGOPE}) and  (\ref{GGOPE}) are equivalent to the
$W_3$
algebra relations
{\footnotesize
	\begin{eqnarray}
	\label{comLL}
	&&[L_n,L_m] = (n-m)L_{n+m} +\frac{c}{12}(n^3-n)
	\delta_{n+m,0} \, , \\
	\label{comLW}
	&&\left[L_n, W_m\right] =\left(2n-m\right)W_{n+m} \, , \\
	\label{comWW}
	&&\left[W_n, W_m\right] =\alpha(n,m) L_{n+m}+\frac{16 (n-m)}{22+5c
	}\Lambda_{n+m}+\frac{c}{360}(n^2-4)(n^2-1)n\delta_{n+m,0},
	\, \,\,\,\,
	\end{eqnarray}
}
where 
\bea
&&\alpha(n,m)=(n-m)\left(\frac{1}{15}(n+m+2)(n+m+3)-\frac{1}{6}(n+2)(m+2)\right)\,,
\nonumber\\
&&\Lambda_{n}=d_nL_n+\sum_{-\infty}^{+\infty}:\!\! L_mL_{n-m}\!\!: \,.
\eea
Here $::$ means normal ordering (i.e operators with smaller index come first) and
\bea
d_{2m}=\frac{1}{5}\left(1-m^2\right)\,,
\quad
d_{2m-1}=\frac{1}{5}\left(1+m\right)\left(2-m\right)\,.
\eea
The central charge of Virasoro algebra in $A_2$-Toda CFT is given by
\bea 
\label{Todac}
c=2+24 q^2\,,
\quad 
q=b+\frac{1}{b}\,,
\eea 
where $b$ is the (dimensionless) Toda coupling. In what follows it would
be convenient to represent the roots, weights and Cartan elements of $A_{2}$ algebra
as $3$-component vectors (endowed with usual Kronecker scalar product) subject
to the condition that sum of components is zero.  In this notation the highest weights
of  fundamental and anti-fundamental representations take the form  
\bea
\label{omega1_2}
\omega_1=\left(
\begin{array}{r}
	2/3\\
	-1/3\\
	-1/3	
\end{array}
\right)\,,
\qquad 
\omega_2=\left(
\begin{array}{r}
	1/3\\
	1/3\\
	-2/3	
\end{array}
\right) 
\,.
\eea  
Furthermore the  weights of  fundamental representation are
\bea 
h_1=\left(
\begin{array}{r}
	2/3\\
	-1/3\\
	-1/3	
\end{array}
\right);\quad h_2=\left(
\begin{array}{r}
	-1/3\\
	2/3\\
	-1/3	
\end{array}
\right) \,;\quad h_3=\left(
\begin{array}{r}
	-1/3\\
	-1/3\\
	2/3	
\end{array}
\right)
\,.
\eea 
 The  Weyl vector  $e_0$ is half sum of positive roots or, alternatively 
 the sum of highest weights of all fundamental representations  
 \bea 
 e_0=\omega_1+\omega_2\,.
 \eea
The conformal (Virasoro) dimensions and $W$-weights of the exponential fields with 
charge $\alpha$ are given by
\bea 
\label{Todadim} 
&&\Delta(\alpha)=\frac{\alpha \cdot (2 Q-\alpha)}{2}\,,\\
\label{Todaw}
&&w(\alpha)=\frac{\sqrt{6} b i}{\sqrt{\left(3 b^2+5\right) \left(5 b^2+3\right)}}\,\prod_{i=1}^3\left(\left(\alpha-
Q\right)\cdot h_i\right)\,,
\eea
where 
\bea
\nonumber
Q=qe_0\,.
\eea
The conjugate charge $\alpha^*$ is defined through conditions
\bea
\label{conj_charge} 
\alpha \cdot \omega_i=\alpha^*\cdot \omega_{3-i}\,;
\quad
i=1,2.
\eea 
If represented as a three component vector, the conjugation amounts to reversing the direction
and permuting the first and third components. 

To pass from the Toda theory to the  minimal models $A_2^{(p)}$, $p=4,5,6,\cdots $, one specifies 
the parameter $b$ as
\bea 
b=i\sqrt{\frac{p}{p+1}}\,.
\eea
 From (\ref{Todac}) for the central charge we get
\bea
\label{cp}
c_p=2-\frac{24}{p(p+1)}\,.
\eea
Furthermore, all primary fields of the minimal models are doubly-degenerated. 
The allowed set of charges is listed below:
{\footnotesize 
\bea
\label{lambdannpmmp}
\alpha \begin{psmallmatrix}
	n & m\\
	n' & m'	
\end{psmallmatrix}
=\frac{i\left(((n-1)(p+1)+(1-m)p)\omega_1
	+((n'-1)(p+1)+(1-m')p)\omega_2 \right)}{\sqrt{p(p+1)}}\,,
\eea 
}
where
$n$, $n'$, $m$, $m'$ are positive integers subject to  constraints
\[n+n'\le p-1, \qquad  m+m'\le p\,.\]
According to (\ref{conj_charge}) the charge of conjugate field is given by
\bea
\label{conj_minimnl} 
\alpha^*
\begin{psmallmatrix}
	n & m\\
	n' & m'	
\end{psmallmatrix}
=\alpha
\begin{psmallmatrix}
	n' & m'\\
	n & m	
\end{psmallmatrix}\,.
\eea
In view of (\ref{Todadim}) the conformal 
dimensions are given explicitly by 
	{\footnotesize
			\bea
			\label{dim}
		\Delta
		\begin{psmallmatrix}
			n & m\\
			n' & m'	
		\end{psmallmatrix}= 	
	\frac{((p+1) (n-n')-p (m-m'))^2+3 ((p+1) (n+n')-p (m+m'))^2-12}{12 p (p+1)}
	\,.
		\eea 
	}
For later reference let us also present here the explicit expression of corresponding $w$ weights 
derived from (\ref{Todaw})
{\footnotesize
\bea 
\label{w3weight}
w\begin{psmallmatrix} n & m\\  n' & m'	\end{psmallmatrix}
&=&\sqrt{\frac{2}{3}}((p+1) (n'-n)-p (m'-m))\times\\
&\times&\frac{  ((p+1) (n+2 n')-p (m+2 m')) ((p+1) (2 n+n')-p (2 m+m'))}{9 p (p+1) \sqrt{(2 p+5) (2 p-3)}}\,.\nonumber
\eea 
}
One can check that the  transformation $\tau$ acting  on charges as
\bea
\tau \circo   \alpha \begin{psmallmatrix} n & m\\  n' & m'	\end{psmallmatrix}:=
\alpha\begin{psmallmatrix} n' & m'\\p-n -n' & p+1-m-m'	\end{psmallmatrix}
\eea
leaves both dimensions and $W_3$ weights  intact.
$\tau$ generates a
$\mathbb{Z}_3$ group $\tau^3=1$. Thus the fields with 
$\tau$ related charges get naturally identified.
\subsection{RG flow between  $A_2^{(p)}$ and $A_2^{(p-1)}$}
In what follows a special role is played by the primary  field $\varphi (x)$ characterized by charge 
\bea
\alpha
\begin{psmallmatrix}
	1 & 2\\
	1 &2	
\end{psmallmatrix}
 =-b e_0\,.
\eea 
This is a relevant field with  conformal dimension 
\bea 
\label{int_field_dim}
\Delta :=\Delta \begin{psmallmatrix}	1 & 2\\	1 &2	\end{psmallmatrix}=
\frac{p-2}{p+1}\equiv 1-\epsilon<1 \,,
\quad
\epsilon=\frac{3}{p+1}.
\eea
 Notice also that the  $\varphi$ is $w$-neutral:
\bea 
\label{wint}
w\begin{psmallmatrix}	1 & 2\\	1 &2	\end{psmallmatrix}=0\,.
\eea
Consider the family generated from this field  by multiple application of OPE. It appears   
that $\varphi$ is the only member of this family  (besides the identity operator)  which is 
relevant. This is important since it allows one   to construct a consistent perturbed CFT with a single coupling constant:
\bea
A=A_{CFT}+g \int \varphi (x)d^2x \,.
\eea 
At large values of $p$, which is the same as  $\epsilon \ll 1$, the perturbing field is only slightly 
relevant and the conformal perturbation theory becomes applicable along a large portion of 
RG flow. In the case of positive  coupling $g >0$ this flow has been investigated 
in \cite{lukyanov1991additional} and it was shown that in the infrared our initial theory $A_2^{(p)}$ flows to 
$A_2^{(p-1)}$. Our aim in this paper is to investigate this RG trajectory in more 
details, in particular investigating the UV/IR mixing matrices for some families of fields. 
One    non-trivial parts of our work 
was computation of some structure constants of OPE, which where not available explicitly  in  literature until now.
For the diagonal structure constants we obtained:  
\begin{scriptsize}
	\bea 
	\label{Cdiag} 
	C^{\begin{psmallmatrix}n & m\\n' & m'	\end{psmallmatrix}}
	_{{\begin{psmallmatrix}1 & 2\\1 & 2		\end{psmallmatrix}},{\begin{psmallmatrix}n & m\\n' & m'		\end{psmallmatrix}}}&=&-\sqrt{\frac{\gamma (3-4 \rho ) \gamma (2-2 \rho )}{\gamma (2-3 \rho )
			\gamma (-2+3 \rho )}}\,\,\frac{ \gamma (\rho )
		\Gamma (2-3 \rho )^2}{ \Gamma (1-\rho )^2}\\
	&\times&\left(\frac{\sin\left( \pi  \left(A_1-2 \rho \right)\right) \gamma \left(\rho+A_1 \right)
		\gamma \left(\rho+B_1 \right)}{\sin \left(\pi  A_1\right) \gamma \left(2-2\rho+A_1\right) 
		\gamma \left(1+B_1\right)}\,
	{{}_{3}F_{2}}\left({1-\rho ,\rho+A_1,1-\rho+B_1
		\atop 1 +B_1, 2 - 2 \rho +A_1};1\right)^2\right.\nonumber\\
	&+&\frac{\sin\left(\pi  \left(A_2-2 \rho \right)\right)
		\gamma \left(\rho -A_1\right) \gamma \left(\rho+A_2 \right)
	}{\sin \left(\pi  A_2\right)
		\gamma \left(2-2\rho+A_2\right)\gamma \left(1-A_1\right)}\,
	{{}_{3}F_{2}}\left({1-\rho ,\rho+A_2 ,1-\rho+B_2
		\atop 1 + B_2, 2 - 2\rho +A_2 };1\right)^2\nonumber \\
	&+&\left.\frac{\sin \left(\pi  \left(A_2+2 \rho \right)\right)\gamma 
		\left(\rho -A_2\right) \gamma \left(\rho -B_1\right)}
	{\sin \left(\pi  A_2\right) \gamma \left(2-2\rho-A_2\right) \gamma \left(1-B_1\right)}\,
	{{}_{3}F_{2}}\left({1-\rho ,\rho+A_3 ,1-\rho+B_3
		\atop 1+B_3,2-2\rho+A_3};1\right)^2\right),\nonumber
	\eea  
\end{scriptsize}
where
\[\gamma(x)\equiv \frac{\Gamma(x)}{\Gamma(1-x)}\,, \quad \rho\equiv p/(p+1)\]
 and
\bea 
&&A_1=-n'+m' \rho\,,\quad  A_2=n+n'-(m+m')\rho \,, \quad  A_3=-A_2;\nonumber\\
&&B_1=n-m \rho\,, \quad  B_2=n'-m' \rho\,, \quad B_3=-B_1\,.
\eea 
The remaining structure constants including the perturbing field $\varphi(x)$ are much 
simpler
\begin{scriptsize}
	\bea 
	\label{offdiag1}
	C^{\begin{psmallmatrix}n & m+1\\n' & m'+1		\end{psmallmatrix}}
	_{{\begin{psmallmatrix}1 & 2\\1 & 2		\end{psmallmatrix}},{\begin{psmallmatrix}n & m\\n' & m'		\end{psmallmatrix}}}
	&=&\frac{(2-3 \rho )\gamma(\rho ) }{n+n'-(m+m'+1)\rho}\times\\
	&\times& \sqrt{\frac{\gamma (3-4 \rho ) \gamma (2-2 \rho ) \gamma (n-m \rho ) \gamma (n'-m' \rho )
			\gamma (n+n'-(m+m')\rho)}{\gamma (1+n-(m+1) \rho) 
			\gamma (1+n'-(m'+1) \rho) \gamma (1+n+n'-(2+m+m')\rho)}}\,,\nonumber\\
	\label{offdiag2}
	C^{\begin{psmallmatrix}n & m-1\\n' & m'+2		\end{psmallmatrix}}
	_{{\begin{psmallmatrix}1 & 2\\1 & 2		\end{psmallmatrix}},{\begin{psmallmatrix}n & m\\n' & m'		\end{psmallmatrix}}}
	&=&\frac{(2-3 \rho )\gamma (\rho )}{n'-(m'+1) \rho}\times\\
	&\times& \sqrt{\frac{ \gamma (3-4 \rho ) \gamma (2-2 \rho ) 
			\gamma (n'-m'\rho ) \gamma (n+\rho (1-m)) \gamma (n+n'-(m+m')\rho)}
		{\gamma (1+n-m \rho) \gamma (1 +n'-(m'+2) \rho) \gamma (1+n+n'-\rho  (m+m'+1))}}\,.\nonumber
	\eea
\end{scriptsize}
For reference let us quot here also the structure constants 
\begin{small}
	\bea
	\label{str_simple1} 
	&&
	C^{\begin{psmallmatrix}n & m+1\\n' & m'		\end{psmallmatrix}}
	_{{\begin{psmallmatrix}1 & 2\\1 & 2		\end{psmallmatrix}},{\begin{psmallmatrix}n & m\\n' & m'		\end{psmallmatrix}}}=
	\sqrt{\frac{\gamma (n+n'-(m+m')\rho) 
			\gamma (-n-n'+(1+m+m')\rho)}{\gamma (1-\rho) 
			\gamma(-2+3 \rho) \gamma (1-n+m \rho) \gamma (1+n-(1+m)\rho)}}\,,\quad\\
	\label{str_simple2}
	&&
	C^{\begin{psmallmatrix}n & m-1\\n' & m'+1		\end{psmallmatrix}}
	_{{\begin{psmallmatrix}1 & 2\\1 & 2		\end{psmallmatrix}},{\begin{psmallmatrix}n & m\\n' & m'		\end{psmallmatrix}}}=
	\sqrt{\frac{\gamma (3-3 \rho ) \gamma (n+(1-m)\rho) 
			\gamma (n'-m'\rho )}{\gamma (1-\rho ) \gamma (1+n-m \rho) 
			\gamma (1+n'-(1+m') \rho)}}\,,\quad\\
	\label{str_simple3}
	&&
	C^{\begin{psmallmatrix}n & m\\n' & m'-1		\end{psmallmatrix}}
	_{{\begin{psmallmatrix}1 & 2\\1 & 2		\end{psmallmatrix}},{\begin{psmallmatrix}n & m\\n' & m'		\end{psmallmatrix}}}=
	\sqrt{\frac{\gamma (3-3 \rho ) \gamma (n'+(1-m') \rho) 
			\gamma (n+n'-(m+m'-1)\rho)}{\gamma (1-\rho ) \gamma (1+n'-m' \rho) 
			\gamma (1+n+n'-(m+m')\rho)}}\,.\quad
	\eea 
\end{small}
The three point function of the field $\varphi$ 
corresponds to the special choice $\alpha=-b(\omega_1+\omega_2)$ in (\ref{Cdiag}). 
The formula gets simplified drastically. The final result reads  (see \cite{lukyanov1991additional})
\bea 
C_{\varphi}\equiv C^{\begin{psmallmatrix}1 & 2\\1 & 2		\end{psmallmatrix}}
_{{\begin{psmallmatrix}1 & 2\\1 & 2		\end{psmallmatrix}},{\begin{psmallmatrix}1 & 2\\1 & 2		\end{psmallmatrix}}} =\frac{2 (4-5 \rho )^2}{(3 \rho -2) (4 \rho -3)}
\frac{\gamma^2\!\left(2-\frac{3 \rho }{2}\right)
	\sqrt{\gamma (4-4 \rho ) \gamma (2-2 \rho)}}{
	\gamma\! \left(1-\frac{\rho }{2}\right) \gamma\! \left(3-\frac{5 \rho }{2}\right) 
	\gamma (3-3 \rho )}\,.\quad 
\eea 
In the limit when $p\gg 1 $ we get
\bea 
&& C_{\varphi}=\frac{3\sqrt{2}}{2}-\frac{3\sqrt{2}}{2}\, \epsilon -\frac{4\sqrt{2}}{3}\, \epsilon ^2
+O(\epsilon^3)\,,
\\
&& \beta (g)=\epsilon\, g  -\frac{\pi }{2}\, C_\varphi\, g^2+O(g^3)\,.
\eea 
Thus at 
\bea 
\label{g}
g=g_*=\frac{2 \sqrt{2}\, \epsilon }{3 \pi }+O\left(\epsilon^2\right)
\eea
the beta-function vanishes and we get an infrared fixed point. The shift of  central charge is given by \cite{Zamolodchikov:1987ti}
\bea
c_p-c_*=12\pi^2\int_{0}^{g_*}\beta(g)dg=\frac{16}{9}\,\epsilon^3+O(\epsilon^4)\,.
\eea 
On the other hand from (\ref{cp})
\bea 
c_p-c_{p-1}=\frac{48}{p (p^2-1)}=\frac{16}{9}\,\epsilon^3+O(\epsilon^4)\,.
\eea 
This strongly supports the identification of  IR fixed point with $A_2^{(p-1)}$ as proposed in \cite{lukyanov1991additional}. 
Later we will give many more evidence in supporting and detailing such identification.

It is well known that the slope of  beta function at a fixed point is directly related to the anomalous dimension 
of the perturbing field. In our case
\bea 
\Delta_*=1-\left.\frac{d\beta}{dg}\right|_{g=g_*}=1+\epsilon+O(\epsilon^2 )\,.
\eea 
As expected the perturbing slightly relevant field $\varphi $ at UV becomes slightly irrelevant at IR.  
Remind that the $w$ weight of $\varphi$ is zero. It is possible to show that 
this $w$ weight doesn't get perturbative corrections at the IR point so that also $w^{IR}=0$.   
Examining (\ref{dim}) and (\ref{w3weight}) we see that the only primary field of $A_2^{(p-1)}$ with charge
\bea 
\alpha
\begin{psmallmatrix}
	2 & 1\\
	2 & 1
\end{psmallmatrix}
=-b^{-1}(\omega_1+\omega_2)
\eea
has vanishing $w$ weight and the  desired dimension
\bea
\left.\Delta
\begin{psmallmatrix}
	2 & 1\\
	2 & 1	
\end{psmallmatrix}
\right|_{p\rightarrow p-1} =\frac{p+2}{p-1}=1+\epsilon+O(\epsilon^2 )\,.
\eea
So that  this field should be  identified with the perturbing field at the IR fixed point.
Another evidence for this conclusion comes from  D.~Gaiotto's RG domain-wall approach 
discussed in section \ref{RGdomaiWall}.  
\section{Matrix of anomalous dimensions}
\label{Matrix_of_anomalous}
According to \cite{Zamolodchikov:1987ti} the  matrix of anomalous dimensions in leading order of coupling constant is given by
\bea
\label{anm_dim_mat}
\Gamma_{\alpha\beta}=\Delta_{\alpha}\delta_{\alpha\beta}+\pi g C_{-be_0,\beta}^{\alpha} +O(g^2)\,.
\eea 
In this section for the first three  smallest RG invariant classes we are going to construct this matrix explicitly.
\subsection{Matrix of anomalous dimension for the first class}
It follows from the structure of OPE that each field
 \bea
 \label{phi0UV}
 \phi\begin{psmallmatrix}n & n\\n' & n'\end{psmallmatrix}
 \eea
 by itself
is RG invariant. From (\ref{Cdiag}) by using following  small $\epsilon$ expansions of generalized hypergeometric functions  of unit argument 
{\footnotesize
\bea
&&	\, _3F_2\left(\frac{\epsilon }{3},1-\frac{\epsilon }{3}(1+n'),\frac{ \epsilon }{3}(1+n);1+\frac{n \epsilon }{3},
\frac{ \epsilon }{3}(2-n');1\right)=\frac{n'-n-3}{n'-2}+O(\epsilon),
\\
&&	\, _3F_2\left(\frac{\epsilon }{3},1-\frac{\epsilon }{3}(1-n-n'),\frac{\epsilon }{3}(1+n');1+\frac{n' \epsilon }{3},\frac{\epsilon }{3}(2+n+n');1\right)= \frac{n+2 n'+3}{n+n'+2}+O(\epsilon),
\\
&&	\, _3F_2\left(\frac{\epsilon }{3},1-\frac{\epsilon }{3}(1+n+n'),\frac{\epsilon }{3}(1-n);1-\frac{n \epsilon }{3},\frac{ \epsilon }{3}(2-n-n');1\right)= \frac{2 n+n'-3}{n+n'-2}+O(\epsilon),
\eea
}
we  obtain
\bea
\label{C0}
	C^{\begin{psmallmatrix}n & n\\n' & n'		\end{psmallmatrix}}
	_{\begin{psmallmatrix}1 & 2\\1 &2\end{psmallmatrix},{\begin{psmallmatrix}n & n\\n' & n'		\end{psmallmatrix}}}=
	\frac{\epsilon ^2 \left(n^2+n n'+n'^2-3\right)}{27 \sqrt{2}}+O\left(\epsilon ^3\right)\,.
\eea
From (\ref{anm_dim_mat}) and  (\ref{g}) at the fixed point for the anomalous dimension we immediately get
\bea
\label{Gamma0}
\Gamma=\frac{1}{27} \epsilon ^2 \left(n^2+n n'+n'^2-3\right)+\frac{1}{27} \epsilon ^3 \left(n^2+n n'+n'^2-3\right)+O\left(\epsilon ^4\right)\,.
\eea
From (\ref{dim}) it is straightforward to see that 
\bea
\Gamma = \Delta^{(p-1)}\begin{psmallmatrix}n & n\\n' & n'	\end{psmallmatrix}
+O\left(\epsilon ^4\right)\,.
\eea
So we conclude that the UV field  $\phi^{(p)}\begin{psmallmatrix}n & n\\n' & n'\end{psmallmatrix}$ 
flows to  
\bea
\label{phi0IR}
\phi^{(p-1)}\begin{psmallmatrix}n & n\\n' & n'\end{psmallmatrix}
\eea
 in the infrared limit.
\subsection{Matrix of anomalous dimensions for the second class}
The second class of RG invariant sets is less trivial.  Each set  contains three primary fields:  
\bea
\label{secund_set_UV}
\phi_1:=\phi\begin{psmallmatrix}n & n+1\\n' & n'	\end{psmallmatrix}\,,\quad
\phi_2:=\phi\begin{psmallmatrix}n & n\\n' & n'-1	\end{psmallmatrix}\,,\quad
\phi_3:=\phi\begin{psmallmatrix}n & n-1\\n' & n'+1\end{psmallmatrix}\,.
\eea
According to (\ref{dim}) their  Virasoro  dimensions for small $\epsilon$  are
\bea
\label{2setUVdim}
&&\Delta_1=\frac{1}{3}-\frac{\epsilon}{9}   (2 n+n'+1)+O\left(\epsilon ^2\right)\,,\\
&&\Delta_2=\frac{1}{3}+\frac{\epsilon }{9}  (n+2 n'-1)+O\left(\epsilon ^2\right)\,,\\
&&\Delta_3=\frac{1}{3}+\frac{\epsilon }{9}  (n-n'-1)+O\left(\epsilon ^2\right)\,.
\eea
The relevant leading order diagonal structure constants are derived  from (\ref{Cdiag}) and (\ref{hyp_sec_set}). Here are the final results:
\bea
\label{st_con11}
&&
C^{\begin{psmallmatrix}n & n+1\\n' & n'	\end{psmallmatrix}}
_{\begin{psmallmatrix}1 & 2\\1 &2\end{psmallmatrix},\begin{psmallmatrix}n & n+1\\n' & n'	\end{psmallmatrix}}
=\frac{2 n (n+n'+3)+3 (n'+1)}{6 \sqrt{2} n (n+n')}+O(\epsilon)\,,\\
\label{st_con22}
&&
C^{\begin{psmallmatrix}n & n\\n' & n'-1	\end{psmallmatrix}}
_{\begin{psmallmatrix}1 & 2\\1 &2\end{psmallmatrix},\begin{psmallmatrix}n & n\\n' & n'-1	\end{psmallmatrix}}
=\frac{n (2 n'-3)+2n'( n'-3)+3}{6 \sqrt{2} n' (n+n')}+O(\epsilon)\,,
\\
\label{st_con33}
&&
C^{\begin{psmallmatrix}n & n-1\\n' & n'+1\end{psmallmatrix}}
_{\begin{psmallmatrix}1 & 2\\1 &2\end{psmallmatrix},\begin{psmallmatrix}n & n-1\\n' & n'+1\end{psmallmatrix}}=
\frac{2 n n'+3 n-3 n'-3}{6 \sqrt{2} n n'}+O(\epsilon)\,.
\eea
The derivation of  off diagonal components is easier. From (\ref{offdiag1}) for small $\epsilon$ we have
\bea
C^{\begin{psmallmatrix}n & n+1\\n' & n'	\end{psmallmatrix}}
_{\begin{psmallmatrix}1 & 2\\1 &2\end{psmallmatrix},\begin{psmallmatrix}n & n\\n' & n'-1	\end{psmallmatrix}}
=\frac{1}{n+n'}\sqrt{\frac{(n+1) (n'-1) \left((n+n')^2-1\right)}{8 nn'}}+O(\epsilon)\,.
\eea
From (\ref{offdiag2})  and the symmetry of structure constants\footnote{
	Remind that $*$ is the conjugation operation  defined in (\ref{conj_minimnl}).}
\bea
 C_{\gamma \beta}^\alpha=C_{\gamma \alpha^{*}}^{\beta^{*}}=C_{\gamma^{*} \alpha}^{\beta}
\eea
for small $\epsilon$ we find 
\bea
C^{\begin{psmallmatrix}n & n+1\\n' & n'	\end{psmallmatrix}}
_{\begin{psmallmatrix}1 & 2\\1 &2\end{psmallmatrix},\begin{psmallmatrix}n & n-1\\n' & n'+1\end{psmallmatrix}}
=\frac{1}{n}\sqrt{\frac{\left(n^2-1\right) (n'+1) (n+n'+1)}{8 n' (n+n')}}+O(\epsilon)\,.
\eea
Similarly
\bea
C^{\begin{psmallmatrix}n & n-1\\n' & n'+1\end{psmallmatrix}}
_{\begin{psmallmatrix}1 & 2\\1 &2\end{psmallmatrix},\begin{psmallmatrix}n & n\\n' & n'-1	\end{psmallmatrix}}
=\frac{1}{n'}\sqrt{\frac{(n-1) \left(n'^2-1\right) (n+n'-1)}{8 n (n+n')}}+O(\epsilon)\,.
\eea
Taking into account these results for the   matrix of anomalous dimensions we get
\begin{small}
	\bea
	\label{Gamma2set}
	&&\Gamma_{11}=\frac{1}{3}-\frac{ \epsilon }{9} \left(1+2 n+n'-\frac{2 n (n+n'+3)+3 (n'+1)}{n (n+n')}\right)+O(\epsilon^2)\,,
	\\
	&&\Gamma_{22}=\frac{1}{3}-\frac{ \epsilon }{9}   \left(1-n-2 n'-\frac{n (2 n'-3)+2 (n'-3) n'+3}{n' (n+n')}\right)+O(\epsilon^2)\,,
	\\
	&&\Gamma_{33}=\frac{1}{3}+\frac{\epsilon}{9}   \left(1+n-n'-\frac{3 (n'+1)}{n n'}+\frac{3}{n'}\right)+O(\epsilon^2)\,,
	\\
	&&\Gamma_{12}=\Gamma_{21}=\frac{\epsilon }{3 (n+n')}\sqrt{\frac{(n+1) (n'-1) \left((n+n')^2-1\right)}{n n'}}+O(\epsilon^2)\,,
	\\
	&&\Gamma_{13}=\Gamma_{31}=\frac{\epsilon }{3 n}\sqrt{\frac{\left(n^2-1\right) (n'+1) (n+n'+1)}{n' (n+n')}}+O(\epsilon^2)\,,
	\\
	&&\Gamma_{23}=\Gamma_{32}=\frac{\epsilon }{3 n'}\sqrt{\frac{(n-1) \left(n'^2-1\right) (n+n'-1)}{n (n+n')}}+O(\epsilon^2)\,.
	\eea
\end{small}
The eigenvalues of this matrix are 
\bea
&&\Delta^{IR}_1=\frac{1}{3}+\frac{\epsilon }{9} (1-2 n-n')+O(\epsilon)\,,
\\
&&\Delta^{IR}_2=\frac{1}{3}+\frac{\epsilon }{9} (1+n+2 n')+O(\epsilon)\,,
\\
&&\Delta^{IR}_3=\frac{1}{3}+\frac{\epsilon }{9} (1+n-n')+O(\epsilon)\,.
\eea
It is not difficult to identify the primary fields of $A_2^{(p-1)}$  
which have such  dimensions with appropriate accuracy
\begin{small}
	\bea
	\label{secund_set_IR}
	\phi_1^{IR}:=\phi^{(p-1)}\begin{psmallmatrix}n-1 & n\\n' & n'	\end{psmallmatrix}\,,\quad
	\phi_2^{IR}:=\phi^{(p-1)}\begin{psmallmatrix}n & n\\n'+1 & n'	\end{psmallmatrix} \,,\quad
	\phi_3^{IR}:=\phi^{(p-1)}\begin{psmallmatrix}n +1& n\\n'-1 & n'	\end{psmallmatrix}\,.
	\eea
\end{small}
To specify the  combinations of our  initial fields $\phi_i$, $i=1,2,3$
which flow to (\ref{secund_set_IR}) one should calculate  
the orthogonal matrix diagonalizing $\Gamma$ 
\bea
(R^T \Gamma R)_{ij}=\Delta^{IR}_{i}\delta_{ij}\,.
\eea
It is straightforward to get convinced that this mixing matrix is given explicitly by 
\begin{scriptsize}
	\bea
	\label{set_two_R}
	R=
	\left(
	\begin{array}{ccc}
		\frac{1}{n (n+n')}\sqrt{\left(n^2-1\right) \left((n+n')^2-1\right)}
		 & \frac{1}{n+n'}\sqrt{\frac{(n+1) (n'+1)}{n n'}}
		 & \frac{}{n}\sqrt{\frac{(n'-1) (n+n'+1)}{n' (n+n')}} \\
		-\frac{1}{n+n'} \sqrt{\frac{(n-1) (n'-1)}{n n'}}
		& \frac{1}{n' (n+n')}\sqrt{\left(n'^2-1\right) \left((n+n')^2-1\right)}
		 & -\frac{1}{n'}\sqrt{\frac{(n+1) (n+n'-1)}{n (n+n')}} \\
		-\frac{1}{n}\sqrt{\frac{(n'+1) (n+n'-1)}{n' (n+n')}}
		 & \frac{1}{n'} \sqrt{\frac{(n-1) (n+n'+1)}{n (n+n')}}
		 & \frac{1}{n n'} \sqrt{\left(n^2-1\right) \left(n'^2-1\right)}\\
	\end{array}
	\right)\,.
	\eea
\end{scriptsize}
Thus we have established that
\bea
\phi^{IR}_i(x)=\sum_{j=1}^{3}\phi_j(x)R_{ji}\,.
\eea
\subsection{Matrix of anomalous dimensions for the third class}
The next class of sets we consider is substantially larger. Each
 set  include   $6$ primaries and $4$ first level descendants:
\bea
\label{uv10}
&&\phi_1=\phi\begin{psmallmatrix}n & n+2\\n' & n'-1	\end{psmallmatrix}\,,\quad
\phi_2=\phi\begin{psmallmatrix}n & n+1\\n' & n'+1	\end{psmallmatrix}\,,\quad
\phi_3=\phi\begin{psmallmatrix}n & n+1\\n' & n'-2	\end{psmallmatrix}\,,\nonumber\\
&&\phi_i=\hat {O}_i\phi\begin{psmallmatrix}n & n\\n' & n'\end{psmallmatrix}\,;
\qquad {\rm for}\,\,\, i=4,5,6,7\,,\\
&&\phi_8=\phi\begin{psmallmatrix}n & n-1\\n' & n'+2	\end{psmallmatrix}\,,\quad
\phi_9=\phi\begin{psmallmatrix}n & n-1\\n' & n'-1	\end{psmallmatrix}\,,\quad
\phi_{10}=\phi\begin{psmallmatrix}n & n-2\\n' & n'+1\end{psmallmatrix}\,,\nonumber
\eea 
where the operators $\hat {O}_i$ are defined as
\bea
&&\hat {O}_4:=L_{-1}\bar{L}_{-1};\quad \hat {O}_5:=L_{-1}(\bar{W}_{-1}-\frac{3w_0}{2\Delta_0}\bar{L}_{-1});\quad
\hat {O}_6:=(W_{-1}-\frac{3w_0}{2\Delta_0}L_{-1})\bar{L}_{-1};\nonumber\\ 
\label{Oi}
&&\hat {O}_7:=(W_{-1}-\frac{3w_0}{2\Delta_0}L_{-1})(\bar{W}_{-1}-\frac{3w_0}{2\Delta_0}\bar{L}_{-1})\,.
\eea 
Here $\Delta_0$  and $w_0$ denote conformal dimension and $W$-weight of the field $\phi\begin{psmallmatrix}n & n\\n' & n'	\end{psmallmatrix}$   (cf. (\ref{w3weight}),(\ref{dim})).
The  specific linear combination of $L_{-1}$ and $W_{-1}$ is chosen so that  the descendant field
\[\left(W_{-1}-\frac{3w_0}{2\Delta_0}L_{-1}\right)\phi\begin{psmallmatrix}n & n\\n' & n'\end{psmallmatrix}\]
is a Virasoro quasi-primary. 
The conformal dimension of fields  $\phi_i$  are
{\footnotesize
	\bea
\nonumber
&\Delta_1=1-\frac{\epsilon}{3} (n+1) +O(\epsilon^2),\,\,
\Delta_2=1-\frac{\epsilon}{3}   (n+n'+1)+O(\epsilon^2 ),\,\,
\Delta_3=1+\frac{\epsilon}{3} (n'-1) +O(\epsilon^2), \qquad\\
&\Delta_{4,5,6,7}=1+O(\epsilon^2 )\,,\qquad
\label{set_3_dim}
\\
&\Delta_8=1-\frac{ \epsilon}{3} (n'+1)+O(\epsilon^2),\,\,
\Delta_9=1+\frac{\epsilon}{3}   (n+n'-1)+O(\epsilon^2),\,\,
\Delta_{10}=1+\frac{\epsilon}{3} (n-1)  +O(\epsilon^2)\,. \quad
\nonumber
\eea
}
We have derived the matrix of  anomalous dimensions using  structure constants given in appendix \ref{st_con_3set}. Here is the final result 
	\bea
	\label{Gamma3set}
	&\Gamma_{11}=1-\epsilon \frac{\left(n^2-5\right) n'^2+n \left(n^2-5\right) n'+n (n+2)}{3 (n+1) n' \left(n'+n\right)}\,,
	\\
	&\Gamma_{12}=\frac{\epsilon }{3 n'}\sqrt{\frac{(n+2) \left(n'+n+2\right) \left(n'^2-1\right)}{(n+1) \left(n'+n+1\right)}}\,,
	\\
	&\Gamma_{13}= \frac{\epsilon}{3 \left(n'+n\right)}\sqrt{\frac{(n+2) \left(n'-2\right) \left(\left(n'+n\right)^2-1\right)}{(n+1) \left(n'-1\right)}}\,,
	\\
	&\Gamma_{14}= \frac{\epsilon(n-2)^2}{2 \left(n^2+n n'+n'^2-3\right)}\sqrt{\frac{(n+2) \left(n'-1\right) \left(n'+n+1\right)}{n n' \left(n'+n\right)}}\,,
	\\
	&\Gamma_{15}= \frac{\epsilon(n-2) \left(2 n'+n\right)}{2  \left(n^2+n n'+n'^2-3\right)}\sqrt{\frac{\left(n^2+n-2\right) \left(n'+1\right) \left(n'+n-1\right)}{3 n (n+1) n' \left(n'+n\right)}}\,,
	\\
	&\Gamma_{16}=\Gamma_{15}\,,
	\\
	&\Gamma_{17}=\epsilon \frac{(n-1) \left(n'+1\right) \left(n'+n-1\right) \left(2 n'+n\right)^2}{6 (n+1) \left(n^2+n n'+n'^2-3\right)}\sqrt{\frac{n+2}{n \left(n'-1\right) n' \left(n'+n\right) \left(n'+n+1\right)}}\,,
	\eea

\bea
&\Gamma_{22}=1+\epsilon \frac{n \left(-n'^3+7 n'+2\right)+n' \left(n'+2\right)-n^3 n'+n^2 \left(1-2 n'^2\right)}{3 n n' \left(n'+n+1\right)}\,,
\\
&	\Gamma_{24}=
\frac{\epsilon \left(n'+n-2\right)^2}{2 \left(n^2+n n'+n'^2-3\right)}
 \sqrt{\frac{(n+1) \left(n'+1\right) \left(n'+n+2\right)}{n n' \left(n'+n\right)}}\,,
\\
&\Gamma_{25}=\frac{\epsilon\left(n'-n\right) \left(n'+n-2\right)}{2 \left(n^2+n n'+n'^2-3\right)}
\sqrt{\frac{(n-1) \left(n'-1\right) \left(n'+n-1\right) \left(n'+n+2\right)}{3 n n' \left(n'+n\right) \left(n'+n+1\right)}}\,,
\\
&\Gamma_{26}=\Gamma_{25}\,,
\\
&\Gamma_{27}=\frac{\epsilon (n-1) \left(n-n'\right)^2 \left(n'-1\right) \left(n'+n-1\right)}{6 \left(n'+n+1\right) \left(n^2+n n'+n'^2-3\right)}
\sqrt{\frac{n'+n+2}{n (n+1) n' \left(n'+1\right) \left(n'+n\right)}}\,,
\\
&\Gamma_{28}=\frac{\epsilon}{3 n}
\sqrt{\frac{\left(n^2-1\right) \left(n'+2\right) \left(n'+n+2\right)}{\left(n'+1\right) \left(n'+n+1\right)}}\,,
\eea
\bea
&\Gamma_{33}=1+\epsilon\frac{n n' \left(n'^2-5\right)+\left(n'-2\right) n'+n^2 \left(n'^2-5\right)}{3 n \left(n'-1\right) \left(n'+n\right)}\,,
\\
&\Gamma_{34}=\frac{\epsilon \left(n'+2\right)^2}{2 \left(n^2+n n'+n'^2-3\right)}
\sqrt{\frac{(n+1) \left(n'-2\right) \left(n'+n-1\right)}{n n' \left(n'+n\right)}}\,,
\\
&\Gamma_{35}=-\frac{\epsilon \left(n'+2\right) \left(n'+2 n\right)}{2  \left(n^2+n n'+n'^2-3\right)}
\sqrt{\frac{(n-1) \left(n'-2\right) \left(n'+1\right) \left(n'+n+1\right)}{3 n \left(n'-1\right) n' \left(n'+n\right)}}\,,
\\
&\Gamma_{36}=\Gamma_{35}\,,
\\
&\Gamma_{37}=\epsilon\frac{(n-1) \left(n'+1\right) \left(n'+n+1\right) \left(n'+2 n\right)^2}{6 \left(n'-1\right) \left(n^2+n n'+n'^2-3\right)}
\sqrt{\frac{n'-2}{n (n+1) n' \left(n'+n\right) \left(n'+n-1\right)}}\,,
\\
&\Gamma_{39}=\frac{\epsilon}{3 n}\sqrt{\frac{\left(n^2-1\right) \left(n'-2\right) \left(n'+n-2\right)}{\left(n'-1\right) \left(n'+n-1\right)}}\,,
\eea
\bea
&\Gamma_{44}=1+\frac{9\epsilon}{2 \left(n^2+n n'+n'^2-3\right)}\,,
\quad
\Gamma_{47}=\frac{9 \epsilon }{2 \left(n^2+n n'+n'^2-3\right)}\,,
\\
&\Gamma_{48}=\frac{\epsilon\left(n'-2\right)^2}{2 \left(n^2+n n'+n'^2-3\right)}
\sqrt{\frac{(n-1) \left(n'+2\right) \left(n'+n+1\right)}{n n' \left(n'+n\right)}}\,,
\\
&\Gamma_{49}=\frac{\epsilon\left(n'+n+2\right)^2}{2 \left(n^2+n n'+n'^2-3\right)}
\sqrt{\frac{(n-1) \left(n'-1\right) \left(n'+n-2\right)}{n n' \left(n'+n\right)}}\,,
\\
&\Gamma_{410}=\frac{\epsilon(n+2)^2}{2 \left(n^2+n n'+n'^2-3\right)}
\sqrt{\frac{(n-2) \left(n'+1\right) \left(n'+n-1\right)}{n n' \left(n'+n\right)}}\,,
\eea
\bea
&\Gamma_{55}=\Gamma_{44}\,,
\qquad
\Gamma_{56}=\Gamma_{47}\,,
\\
&\Gamma_{57}=\frac{\epsilon  (n'-n) (2 n+n') (n+2 n')}{n^2+n n'+n'^2-3}
\sqrt{\frac{3}{\left(n^2-1\right) \left(n'^2-1\right) \left((n+n')^2-1\right)}}\,,
\\
&\Gamma_{58}=-\epsilon\frac{\left(n'-2\right) \left(n'+2 n\right)}{2  \left(n^2+n n'+n'^2-3\right)}
\sqrt{\frac{(n+1) \left(n'+n-1\right) \left(n'^2+n'-2\right)}{3n n' \left(n'+1\right) \left(n'+n\right)}}\,,
\\
&\Gamma_{59}=\epsilon\frac{\left(n'-n\right) \left(n'+n+2\right)}{2  \left(n^2+n n'+n'^2-3\right)}
\sqrt{\frac{(n+1) \left(n'+1\right) \left(n'+n-2\right) \left(n'+n+1\right)}{3 n n' \left(n'+n-1\right) \left(n'+n\right)}}\,,
\\
&\Gamma_{510}=\epsilon \frac{(n+2) \left(2 n'+n\right)}{2  \left(n^2+n n'+n'^2-3\right)}
\sqrt{\frac{(n-2) (n+1) \left(n'-1\right) \left(n'+n+1\right)}{3 (n-1) n n' \left(n'+n\right)}}\,,
\eea
\bea
&\Gamma_{66}=\Gamma_{55}\,,
\quad
\Gamma_{67}=\Gamma_{57}\,, 
\quad
\Gamma_{68}=\Gamma_{58}\,,
\quad
\Gamma_{69}=\Gamma_{59}\,,
\quad
\Gamma_{610}=\Gamma_{510}\,,
\eea
\bea
&\Gamma_{77}=1+\epsilon 
\frac{3 \left(n-n'\right)^2 \left(n'+2 n\right)^2 \left(2 n'+n\right)^2+4 \left(n^2+n n'+n'^2-3\right)^3+9 \left(n^2+n n'+n'^2-3\right)^2}{6 \left(n^2-1\right) \left(n'^2-1\right) \left(n^2+n n'+n'^2-3\right) \left(\left(n'+n\right)^2-1\right)}\,,
\\
&\Gamma_{78}=\epsilon \frac{(n+1) \left(n'-1\right) \left(n'+n-1\right) \left(n'+2 n\right)^2}{6 \left(n'+1\right) \left(n^2+n n'+n'^2-3\right)}
\sqrt{\frac{n'+2}{(n-1) n n' \left(n'+n\right) \left(n'+n+1\right)}}\,,
\\
&\Gamma_{79}=\epsilon \frac{(n+1) \left(n-n'\right)^2 \left(n'+1\right) \left(n'+n+1\right)}{6 \left(n'+n-1\right) \left(n^2+n n'+n'^2-3\right)}
\sqrt{\frac{n'+n-2}{(n-1) n \left(n'-1\right) n' \left(n'+n\right)}}\,,
\\
&\Gamma_{710}=\epsilon \frac{(n+1) \left(n'-1\right) \left(n'+n+1\right) \left(2 n'+n\right)^2}{6 (n-1) \left(n'+1\right) \left(n'+n-1\right) \left(n^2+n n'+n'^2-3\right)}
\sqrt{\frac{(n-2) \left(n'+1\right) \left(n'+n-1\right)}{n n' \left(n'+n\right)}}\,,
\eea
\bea
&\Gamma_{88}=1-\epsilon \frac{n n' \left(n'^2-5\right)+n' \left(n'+2\right)+n^2 \left(n'^2-5\right)}{3 n \left(n'+1\right) \left(n'+n\right)}\,,
\\
&\Gamma_{810}=\frac{\epsilon}{3 \left(n'+n\right)}\sqrt{\frac{(n-2) \left(n'+2\right) \left(\left(n'+n\right)^2-1\right)}{(n-1) \left(n'+1\right)}}\,,
\eea
\bea
&\Gamma_{99}=1+
\epsilon \frac{n \left(n'^3-7 n'+2\right)-\left(n'-2\right) n'+n^3 n'+n^2 \left(2 n'^2-1\right)}{3 n n' \left(n'+n-1\right)}\,,
\\
&\Gamma_{910}=
\frac{\epsilon}{3 n'}
\sqrt{\frac{(n-2) \left(n'+n-2\right) \left(n'^2-1\right)}{(n-1) \left(n'+n-1\right)}}\,,
\eea
\bea
&\Gamma_{1010}=1+\epsilon
\frac{\left(n^2-5\right) n'^2+n \left(n^2-5\right) n'+(n-2) n}{3 (n-1) n' \left(n'+n\right)}\,.
\eea
Remind that  $\Gamma$ is symmetric and the matrix elements not listed above are  equal to $0$.
Its eigenvalues  are
{\footnotesize
\bea
&\Delta_1^{IR}=1+ \frac{\epsilon }{3} (1+n')+O(\epsilon^2)\,,
\,\,
\Delta_2^{IR}=1+\frac{\epsilon}{3}   (1+n+n')+O(\epsilon^2)\,,
\,\,
\Delta_3^{IR}=1+\frac{\epsilon}{3}\left(1-n\right)+O(\epsilon^2)  \,,
\qquad
\nonumber
\\
&\Delta_{4,5,6,7}^{IR}=1+O(\epsilon^2)\,,
\qquad
\\
&\Delta_8^{IR}=1+\frac{\epsilon }{3} (1+n)+O(\epsilon^2) \,,
\,\,
\Delta_9^{IR}=1+\frac{ \epsilon }{3} (1-n-n')+O(\epsilon^2)\,,
\,\,
\Delta_{10}^{IR}=1+\frac{ \epsilon}{3}\left(1-n'\right)+O(\epsilon^2) \,.
\qquad
\nonumber
\eea
}
Using (\ref{dim}) we see that the $A_2^{(p)}$ fields given in   (\ref{uv10}) flow to 
\bea
\label{ir10}
&&\phi^{(p-1)}\begin{psmallmatrix}n-1 & n\\n'+2 & n'  \end{psmallmatrix}\,,\quad
\phi^{(p-1)}\begin{psmallmatrix}n+1 & n\\n'+1 & n'	\end{psmallmatrix}\,,\quad
\phi^{(p-1)}\begin{psmallmatrix}n-2 & n\\n'+1 & n'	\end{psmallmatrix}\,,\nonumber\\
&&\phi^{(p-1)}_i=\hat {O}_i\phi\begin{psmallmatrix}n & n\\n' & n'\end{psmallmatrix}\,;
\qquad {\rm for}\,\,\, i=4,5,6,7\,,\\
&&\phi^{(p-1)}\begin{psmallmatrix} n+2 & n\\n'-1 & n'	\end{psmallmatrix}\,,\quad
\phi^{(p-1)}\begin{psmallmatrix}n-1 & n\\n'-1 & n'	\end{psmallmatrix}\,,\quad
\phi^{(p-1)}\begin{psmallmatrix}n+1 & n\\n'-2 & n'	\end{psmallmatrix} \nonumber
\eea 
in $A_2^{(p-1)}$.
\section{ Matrix of anomalous $W$-weights}
\label{Wanom}
In perturbed theory $W$ is no longer holomorphic. Using   Ward identity for $W$ current \cite{Fateev:2007ab}
together with (\ref{wint}) and null vector condition   
\[
W_{-2}\varphi(z_i)=\frac{2}{\Delta+1}\partial_{z_i}W_{-1}\varphi(z_i)
\] 
one can verify  the following equality 
{\footnotesize
	\bea
	\label{Wint}
	&\bar{\partial}\langle W(z)e^{-\int g \varphi d^2 x}\rangle=
	\sum_{n=0}^{\infty}\sum_{i=1}^{n}\frac{(-g)^n}{n!}\times
	\\ \nonumber
	&\int
	\bar{\partial}\langle
	\varphi(x_1)\dots 
	\left(
	\frac{1}{(z-x_i)^2}+\frac{2}{(\Delta+1)(z-x_i)}\frac{\partial}{\partial x_i}
	\right)W_{-1}\varphi(x_i)
	\dots\varphi(x_n)\rangle
	d^2x_1\dots d^2x_n \,.
	\eea
}
Next we  use
\bea 
\label{dif_id}
\bar{\partial}(z-x_i)^{-1}=\pi \delta^{(2)}(z-x_i),\quad {\rm hence} \quad
\bar{\partial}(z-x_i)^{-2}=-\pi \delta^{\prime}(z-x_i) 
\eea
and evaluate the integral over $x_i$ in (\ref{Wint}). The result can be conveniently represented as  
\bea
\bar{\partial}\langle W(z)e^{-\int g \varphi d^2 x}\rangle=
-g \pi \frac{1-\Delta}{1+\Delta} \langle  (\partial W_{-1} \varphi(z)) e^{-\int g \varphi d^2 x} \rangle\,,
\eea
which indicates that the following  conservation law is satisfied
\bea 
\label{BardW}
\bar{\partial}W(z,\bar{z})+ \pi g \frac{1-\Delta}{1+\Delta}\partial W_{-1}\varphi(z,\bar{z})
=0\,.
\eea
This identity can be rewritten as
\bea
\label{con_cur_W}
\bar{\partial}(z^2W)+ \pi g \frac{1-\Delta}{1+\Delta}\partial(z^2W_{-1}\varphi)=2 \pi g \frac{1-\Delta}{1+\Delta}zW_{-1}\varphi\,.
\eea   
Certainly this is not a conservation law (due to non zero right hand side). Still a charge  (non conserved)
denoted by $\mathfrak{Q}$ - the analog of $W$ zero mode in perturbed theory  can be defined as
\bea
\label{Def_WanomDim}
\mathfrak{Q}
=\int_{{\cal C}}\left(z^2W(z)\frac{dz}{2\pi i}- \pi g \frac{1-\Delta}{1+\Delta}z^2W_{-1}\varphi(z)\frac{d\bar{z}}{2\pi i}\right)\,.
\eea
Since in 2d QFT it is common to consider radial quantization one can choose  
$\cal C$ to be a circle centered at zero.
Let $\Lambda$ be a region 
 bounded by two circles one of  large radius $R$ and the second with a  very small   radius $\varepsilon$.
Integrating  the lhs of  equation (\ref{con_cur_W}) in presence of a field located at zero
and applying  Stokes' theorem we find
\bea
\left(\oint_{{\cal C}_R}-\oint_{{\cal C}_\varepsilon}\right)
\left(z^2W(z)\frac{dz}{2\pi i}- \pi g \frac{1-\Delta}{1+\Delta}z^2W_{-1}\varphi(z)\frac{d\bar{z}}{2\pi i}\right)\phi_\beta(0)\,.
\nonumber
\eea
Due to smallness of $\varepsilon$ the influence of interaction is irrelevant and the contribution of ${\cal C}_\varepsilon$  is equal to
$w_\beta \phi_\beta$ where $w_\beta$ is the W-weight of $\phi_\beta$ in unperturbed UV theory. Instead  the contribution
of  large contour  ${\cal C}_R$ is just the variation of $\phi_\beta$ by the charge $Q$. Thus we finally arrive at
\bea
\label{Qw_as_int}
\hat{\mathfrak{Q}}(\phi_\alpha)=
w_\alpha\phi_\alpha(0)
+2 \pi g \frac{\epsilon}{2-\epsilon}\int_{\Lambda}zW_{-1}\varphi(z)\phi_\alpha(0)\frac{d\bar{z}dz}{2\pi i}\,.
\eea
   Consider the OPE
\bea
W_{-1}\varphi(z)\phi_\alpha(0)=\sum_{\beta}z^{\Delta_\beta-\Delta-\Delta_\alpha-1}b_\alpha^\beta\phi_\beta+\dots \,.
\eea
It is convenient to represent the structure constants  $b_\alpha^\beta$ as
\bea
\label{bOffDiagonal}
&b^\beta_{\alpha }=\left(1-\frac{\epsilon}{2}\right)a_{ \alpha \beta } C_{\,\,\alpha}^\beta \,.
\eea
It follows from $W$-Ward identities (see  (\ref{Wphi3f})) that
\bea
\label{aOffDiagonal}
a_{\alpha \beta }=\frac{w_{\alpha}-w_{\beta}}{\Delta_{\alpha}-\Delta_{\beta}}
\qquad {\rm if} \qquad \alpha \ne \beta\,.
\eea
The diagonal element  $a_{\alpha \alpha }$ can  not be  determined by Ward identities alone.
Still in cases of our interest we are able to determine it by
investigating four point correlation functions (see appendix \ref{AppStCons}).  
In (\ref{Qw_as_int}) we pass to radial coordinates $d\bar{z} dz=2irdrd\theta$  and the integral over
$r$ is  evaluated from $[0,R]$ where $R$ is  an  infrared scale which for convenience we choose 
$R=1$. The result is
\bea
\label{opeWm1}
\hat{\mathfrak{Q}}(\phi_\alpha)=
w_\alpha\phi_\alpha(0)
+\pi g \epsilon \sum_{\beta}\frac{ a_{\alpha \beta } C^\beta_{\,\,\alpha}}{\Delta_{\beta\alpha}+\epsilon}\phi_\beta(0)\,.
\eea
Let us introduce the renormalized fields (see appendix \ref{ap_phi_lampda} for details)
\bea
\label{phiTilde}
&\tilde{\phi}_\alpha=\phi_\alpha+
\sum_\beta \frac{\pi g C_{\,\,\alpha}^\beta}{\Delta_{\beta\alpha}+\epsilon}\,\phi_\beta\,.
\eea
From (\ref{opeWm1})
{\footnotesize
\bea
\hat{\mathfrak{Q}}\left(\tilde{\phi}_\alpha\right)&=&
w_\alpha\phi_\alpha
+\frac{ \pi g \epsilon a_{\alpha \beta} C^\beta_{\,\,\alpha}}{\Delta_{\beta\alpha}+\epsilon}\phi_\beta+
\frac{\pi g w_\beta C_{\,\,\alpha}^\beta}{\Delta_{\beta\alpha}+\epsilon}\phi_\beta+O(g^2)=
\\ \nonumber
&& w_\alpha\left(\phi_\alpha+\frac{\pi g C_{\,\,\alpha}^\beta}{\Delta_{\beta\alpha}+\epsilon}\phi_\beta\right)
-\frac{\pi g w_\alpha C_{\,\,\alpha}^\beta}{\Delta_{\beta\alpha}+\epsilon}\phi_\beta
+\frac{ \pi g \epsilon a_{\alpha \beta} C^\beta_{\,\,\alpha}}{\Delta_{\beta\alpha}+\epsilon}\phi_\beta+
\frac{\pi g w_\beta C_{\,\,\alpha}^\beta}{\Delta_{\beta\alpha}+\epsilon}\phi_\beta+O(g^2)
\eea
}
or
\bea
\hat{\mathfrak{Q}}(\tilde{\phi}_\alpha)=
w_\alpha \tilde{\phi}_\alpha+\frac{\pi g C_{\,\,\alpha}^\beta}{\Delta_{\beta\alpha}+\epsilon}
\left(w_{\beta\alpha} +\epsilon a_{\alpha \beta} \right)\tilde{\phi}_\beta+O(g^2)\,.
\eea
Using (\ref{aOffDiagonal}) we get 
\bea
\hat{\mathfrak{Q}}(\tilde{\phi}_\alpha)=
w_\alpha \tilde{\phi}_\alpha
+\pi g \sum_{\beta} a_{\alpha \beta}C_{\,\,\alpha}^\beta \tilde{\phi}_\alpha+O(g^2)\,.
\eea
Thus for the matrix element of the $W$-charge $\mathfrak{Q}$ we find 
\bea
\label{QmatEl}
\mathfrak{Q}_{\alpha\beta}=w_{\beta}\delta_{\alpha\beta}
+\pi g a_{\alpha\beta} C_{\,\,\alpha}^\beta+O(g^2)\,.
\eea
\subsection{Matrix of anomalous $W$-weights for the first class}
Here each RG invariant set consists of a single field (\ref{phi0UV}). Its $W$-weight is
\begin{small}
	\bea
	w\begin{psmallmatrix}n & n\\n' & n'\end{psmallmatrix}=
	\frac{\epsilon ^3 (n'-n) (2 n+n') (n+2 n')}{243 \sqrt{6}}+
	\frac{\epsilon ^4 (n'-n) (2 n+n') (n+2 n')}{486 \sqrt{6}}+O(\epsilon^5)\,.
	\eea
\end{small}
For this case the coefficient $a$ entering in (\ref{QmatEl}) is given by eq. (\ref{a_coef}).
Taking into account also   (\ref{C0}) and  (\ref{g})  we obtain
\begin{small}
	\bea
	\label{W0IR}
	\mathfrak{Q}=\frac{\epsilon ^3 (n'-n) (2 n+n') (n+2 n')}{243 \sqrt{6}}+\frac{\epsilon ^4 (n'-n) (2 n+n') (n+2 n')}{162 \sqrt{6}}+O(\epsilon^5)\,,
	\eea
\end{small}
which indeed coincides with $W$-weight of $\phi^{(p-1)}\begin{psmallmatrix}n & n\\n' & n'\end{psmallmatrix}$ with desired accuracy.
\subsection{Matrix of anomalous $W$-weights for the second class}
The  fields of second class  (\ref{secund_set_UV}) in UV limit have the following $W$-dimensions 
\begin{small}
	\bea
	w_1&=&\frac{2}{9 \sqrt{6}}-\frac{\epsilon  }{9 \sqrt{6}}(2 n+n'+1)+O(\epsilon^2)\,,
	\\
	w_2&=&\frac{2}{9 \sqrt{6}}+\frac{\epsilon  }{9 \sqrt{6}}(n+2 n'-1)+O(\epsilon^2)\,,
	\\
	w_3&=&\frac{2}{9 \sqrt{6}}+\frac{\epsilon  }{9 \sqrt{6}}(n-n'-1)+O(\epsilon^2)\,.
	\eea
\end{small}
Comparison with  (\ref{2setUVdim}) ensures that for $\alpha \ne \beta$ 
\[
\frac{w_{\alpha}-w_{\beta}}{\Delta_{\alpha}-\Delta_{\beta}}=\frac{1}{\sqrt{6}}\,.
\]
In appendix \ref{AppStCons}  we show how to derive the  value
\bea
\label{a_alp_alp}
a_{\alpha\alpha}=1/\sqrt{6}\,.
\eea
 Thus (\ref{QmatEl}) 
becomes
\bea
\label{W_anm_dim_mat}
\mathfrak{Q}_{\alpha\beta}=w_{\beta}\delta_{\alpha\beta}+\frac{\pi g}{\sqrt{6}}  C_{\,\, \beta}^{\alpha}+
O(g^2)\,.
\eea
Explicitly
{\footnotesize
	\bea
	&&\mathfrak{Q}_{11}=\frac{2}{9 \sqrt{6}}+\frac{\epsilon}{9 \sqrt{6}}
	\left(\frac{2 n (n+n'+3)+3 (n'+1)}{n (n+n')}-2 n-n'-1\right)+O(\epsilon^2)\,,
	\\
	&&\mathfrak{Q}_{22}=\frac{2}{9 \sqrt{6}}+\frac{\epsilon  }{9 \sqrt{6}}
	\left(\frac{n (2 n'-3)+2 (n'-3) n'+3}{n' (n+n')}+n+2 n'-1\right)+O(\epsilon^2)\,,
	\\
	&&\mathfrak{Q}_{33}=\frac{2}{9 \sqrt{6}}+\frac{\epsilon}{9 \sqrt{6}}
	\left(\frac{3 n-3 (n'+1)}{n n'}+n-n'+1\right)+O(\epsilon^2)\,,
	\\
	&&\mathfrak{Q}_{12}=\frac{\epsilon  }{3 \sqrt{6} (n+n')}
	\sqrt{\frac{(n+1) (n'-1) \left((n+n')^2-1\right)}{n n'}}+O(\epsilon^2)\,,
	\\
	&&\mathfrak{Q}_{13}=\frac{\epsilon  }{3 \sqrt{6} n}
	\sqrt{\frac{\left(n^2-1\right) (n'+1) (n+n'+1)}{n' (n+n')}}+O(\epsilon^2)\,,
	\\
	&&\mathfrak{Q}_{23}=
	\frac{\epsilon  }{3 \sqrt{6} n'}
	\sqrt{\frac{(n-1) \left(n'^2-1\right) (n+n'-1)}{n (n+n')}}+O(\epsilon^2)\,.
	\eea
}
Here are the eigenvalues of this matrix
\bea
&&w^{IR}_1=\frac{2}{9 \sqrt{6}}-\frac{\epsilon  (2 n+n'-1)}{9 \sqrt{6}}+O(\epsilon^2)\,,
\\
&&w^{IR}_2=\frac{2}{9 \sqrt{6}}+\frac{\epsilon  (n+2 n'+1)}{9 \sqrt{6}}+O(\epsilon^2)\,,
\\
&&w^{IR}_3=\frac{2}{9 \sqrt{6}}+\frac{\epsilon  (n-n'+1)}{9 \sqrt{6}}+O(\epsilon^2)\,,
\eea
which coincide with the $w$ dimensions of the fields (\ref{secund_set_IR})
in $A_2^{(p-1)}$ as  expected. 
\section{The RG domain wall}
\label{RGdomaiWall}
In \cite{Gaiotto:2012np} D.~Gaiotto suggests
a candidate for RG domain wall for the
general RG flow between theories (\ref{GenUV}) and (\ref{GenIR}).
Let us briefly recall his construction.
Any  conformal
interface between two CFTs can be alternatively represented as 
a conformal boundary condition  for the direct product of these
 two theories (folding trick). This boundary condition encodes
 the UV/IR map. 
 The construction of appropriate boundary is heavily based on
 coset representation of $W$ minimal models (\ref{cosetrep}).  
 For the direct product theory we get
\begin{eqnarray}
	\label{prodTh}
\mathcal{T}_{UV}\times \mathcal{T}_{IR} \sim
	\frac{ \hat{g}_{m-l} \times \hat{g}_l}{\hat{g}_{m}}
	\times 
	\frac{ \hat{g}_{m}\times \hat{g}_l }{\hat{g}_{m+l}}
	\sim \frac{\hat{g}_{m-l}\times \hat{g}_l \times \hat{g}_l }{\hat{g}_{l+m}} \,.
\end{eqnarray}
Notice the appearance of two identical factors $\hat{g}_l$ in the numerator.
Hence the resulting theory admits a non trivial 
 ${\mathbb{Z}}_2$ symmetry which intertwines these two factors.
Essentially D.~Gaiotto conjecture  boils down to the
statement that  boundary of the theory
\begin{eqnarray}
	\mathcal{T}_B=\frac{\hat{g}_l \times \hat{g}_l \times \hat{g}_{m-l} }{\hat{g}_{l+m}}, \qquad m>l
\end{eqnarray}
acts as a ${\mathbb{Z}}_2$ twisting mirror. 
The basic building block of this boundary state
is defined by the property that  mirror image of
any local field of product theory is identified with
its   ${\mathbb{Z}}_2$ counterpart.
As in standard Cardy construction these 
building blocks should be superposed with
appropriate coefficients in order to get
the ``physical" boundary state. 

These coefficients have been explicitly  identified in \cite{Gaiotto:2012np}
in terms of current algebra modular matrix. The final
expression can be formally displayed as
\begin{equation}
		\label{Gen_boundary}
	|\tilde{B} \rangle = \sum\limits_{s,t}\sqrt{S^{(m-l)}_{1,t}S^{(m+l)}_{1,s}}\sum\limits_d |t,d,d,s;Z_2\rangle \rangle ,
\end{equation}
where the indices $t$, $d$, $s$ refer to the representations
of $\hat{g}_{m-l}$, $\hat{g}_l$, $\hat{g}_{l+m}$ respectively,
 $S^{(k)}_{1,r}$ are the modular matrices of the
$\hat{g}_{k}$ WZNW model and  $|t,d,d,s;Z_2\rangle \rangle $ is above mentioned ${\mathbb{Z}}_2$
image of the Ishibashi state  $|t,d,d,s\rangle \rangle$ of the  product theory.

In what follows we will examine in details the
RG flow between $W_3$ minimal models for our second class. 
\subsection{Current algebra}
\label{Coset construction}
The WZNW models have extended chiral  symmetry  generated by 
 Virasoro  and  ${\rm Ka\breve{c}}$-${\rm Moody}$ algebras.
 The ${\rm Ka\breve{c}-Moody}$ generators  can be considered as
 Laurent coefficients of spin one holomorphic currents.
  In case of $SU(3)$ WZNW models we deal with  the algebra  $\hat{A}_2$. It will be convenient for us 
  to choose generators (inherited from standard $sl(3)$ Cartan-Weyl  basis)  $E^{ij}_n$ and $H_n^i$  with $i,j\in \{1,2,3\}$, $i\ne j$ 
  and   $n\in \mathbb{Z}$ is the loop index. 
\bea
\nonumber
&&\left[E_n^{ij}E_m^{rl}\right]=\delta^{jr}E_{n+m}^{il}-\delta^{li}E_{n+m}^{rj}+kn\delta_{n+m}\delta^{il}\delta^{jr}\,,
\\
\label{KM_alg}
&&\left[E_n^{ij}H_m^{r}\right]=\delta^{jr}E_{n+m}^{ir}-\delta^{ri}E_{n+m}^{rj}\,,
\\
\nonumber
&&\left[H_n^{i}H_m^{j}\right]=k n \delta_{n+m}\left(\delta_{ij}-\frac{1}{3}\right)\,,
\eea
where $k=1,2,3....$ referred as the level specifies a model. It is also assumed that
\[
\sum_{i=1}^3H^i_n=0\,.
\]
The Virasoro generators can be expressed in terms of  current algebra  through the Sugawara construction
\begin{align}
\label{Ln_WZW_gen}
L_n=\frac{1}{2(k+3)}\sum_{l\in Z}\sum_{i,j=1}^3:E_l^{ij}E_{n-l}^{ji}:\,.
\end{align}
The normal ordering prescription  is the standard one: 
The order of generators  should be flipped provided 
the loop index of  first generator is less than that of the second 
one.  

Using (\ref{KM_alg}) and  (\ref{Ln_WZW_gen}) one can get convinced that the 
 Virasoro charge equals to
\begin{eqnarray}
c_k=\frac{8k}{k+3} \,.
\label{KM central charge}
\end{eqnarray}
The primary fields of the theory (denoted as $\phi_{\lambda}$)  are labeled by the
highest weights 
\bea
\label{lambda}
\lambda=(n-1)\omega_1+(n'-1)\omega_2\,,
\eea
where $n$, $n'$ are positive integers and $\omega_1$, $\omega_2$ are the same as in  (\ref{omega1_2}).
The following formula for   conformal dimensions of primary fields can be readily  deduced from (\ref{Ln_WZW_gen})
\bea
h_{k}(\lambda)=\frac{\lambda \cdot (\lambda+2e_0)}{2(k+3)}\,.
\label{WZW_dim}
\eea
The corresponding primary states satisfy the conditions
\bea
&& H^i_0 |\phi_\lambda \rangle=\lambda_{ii}|\phi_\lambda \rangle\,,
\\
&& E_0^{ij}|\phi_\lambda \rangle=0\quad {\rm for} \quad i<j\,,
\\
&& E_n^{ij}|\phi_\lambda \rangle=H_n^{i}|\phi_\lambda \rangle=0 \quad {\rm for} \quad n>0\,.
\eea
To construct the domain wall we need also the explicit form of the $\hat{A}_2$
 modular matrix \cite{Gannon:1992ty} which in notations specified above takes the form
{\footnotesize
		\bea
	S^{(k)}_{\lambda_{n,n'},\lambda_{m,m'}}
	\equiv S^{(k)}_{\begin{psmallmatrix} n\\n'	\end{psmallmatrix}
		\begin{psmallmatrix} m\\m'	\end{psmallmatrix}}
	=\frac{i^3}{(k+3)\sqrt{3}}\sum_{w\in W}\varepsilon(w)     
	{\rm exp}\left[
	-2\pi i\frac{w(\lambda_{m+1,m'+1})\cdot \lambda_{n+1,n'+1}}{k+3}
	\right].
	\eea
}
Here the sum runs over 6 elements of the Weil  group $W$ acting on three component vectors through permutations.
More explicitly
{\footnotesize
\bea
\label{su3_mod_matrix}
S^{(k)}_{\begin{psmallmatrix} n\\n'	\end{psmallmatrix}\begin{psmallmatrix} m\\m'	\end{psmallmatrix}}=
\frac{-i}{\sqrt{3} (k+3)}
\bigg(
e_k\left[2 m' n'+n m'+m n'-m n\right]-e_k\left[2 m' n'+n m'+m n'+2 m n\right]
\\ \nonumber
+e_k\left[-m' n'+n m'+m n'+2 m n\right]-e_k\left[-m' n'-2 n m'+m n'-m n\right]
\\ \nonumber
-e_k\left[-m' n'+n m'-2 m n'-m n\right]+e_k\left[-m' n'-2 n m'-2 m n'-m n\right]
\bigg).
\eea
}
with notation
\bea
e_k[x]\text{:=}e^{-\frac{2 \pi  i x}{3 (k+3)}}\,.
\eea
According  to
\cite{Goddard:1984vk, Goddard:1986ee}
the $W_3$ minimal models can be alternatively
represented  as coset theory 
\bea
\label{W3coset}
\frac{\widehat{su}(3)_k \times \widehat{su}(3)_1}{\widehat{su}(3)_{k+1}}\,,
\eea
where   $\widehat{su}(3)_k$ stands for level $k$ WZNW model
\cite{Knizhnik:1984nr, Zamolodchikov:1986bd} with identification
\bea
\label{p_k_conection}
p=k+3\,.
\eea
 The stress energy tensor of the coset theory (\ref{W3coset}) is given by
\begin{equation}
T_{(su(3)_k\times su(3)_1)/su(3)_{k+1}}=T_{su(3)_k}+T_{su(3)_1}-T_{su(3)_{k+1}} \,. 
\label{diagonalcoset}
\end{equation}
In particular for the central charge we get
\bea
\label{cCoset}
c_{(su(3)_k\times su(3)_1)/su(3)_{k+1}}=c_{su(3)_k}+c_{su(3)_1}-c_{su(3)_{k+1}} \,. 
\eea
Taking into count (\ref{KM central charge}) one easily gets convinced that 
(\ref{cCoset}) coincides with the central charge of $A_2^{(p)}$ minimal model (\ref{cp}) with already mentioned identification
(\ref{p_k_conection}).
\subsection{The IR/UV mixing coefficients through  domain wall approach}
For the $W_3$ minimal models (\ref{prodTh}) becomes
\bea
\label{cos_W3}
\underbrace{\frac{\widehat{su}(3)_{k-1} \times \widehat{su}(3)_1}{\widehat{su}(3)_{k}}}_{{\rm IR : \,\, A_2^{(k+2)}}}
\times
\underbrace{\frac{\widehat{su}(3)_k \times \widehat{su}(3)_1}{\widehat{su}(3)_{k+1}}}_{\rm UV: \,\, A_2^{(k+3)}}
\sim
\frac{\widehat{su}(3)_{k-1} \times \widehat{su}(3)_1 \times \widehat{su}(3)_1}{\widehat{su}(3)_{k+1}}\,.
\eea
The formula for  mixing coefficients  dictated by (\ref{Gen_boundary}) is
\bea
\label{mix_coef_dom}
&\langle \phi^{IR} \begin{psmallmatrix}m_i & n\\m_i' & n'	\end{psmallmatrix}\phi^{UV} \begin{psmallmatrix}n & m_j\\n' & m_j'	\end{psmallmatrix} | RG \rangle=
\frac{\langle \widetilde{ij} | ij \rangle}{ \langle ij | ij \rangle}
\frac{\sqrt{S^{(k-1)}_{\begin{psmallmatrix} 1\\1 \end{psmallmatrix}\begin{psmallmatrix} m_i \\ m_i'	\end{psmallmatrix}} S^{(k+1)}_{\begin{psmallmatrix} 1 \\1\end{psmallmatrix} \begin{psmallmatrix} m_j \\ m_j'	\end{psmallmatrix}}}}
{S^{(k)}_{\begin{psmallmatrix} 1 \\ 1\end{psmallmatrix} \begin{psmallmatrix} n \\ n'	\end{psmallmatrix}}}\,.
\eea
For the second class we set  $\phi^{UV} \begin{psmallmatrix}n & m_j\\n' & m_j'	\end{psmallmatrix}\equiv \phi_j$   from  (\ref{secund_set_UV})
and $\phi^{IR} \begin{psmallmatrix}m_i & n\\m_i' & n'	\end{psmallmatrix}\equiv \phi_i^{IR}$ from (\ref{secund_set_IR}).
$ | ij \rangle$ is the representative of state $ \phi^{IR}_{i}\phi_j | 0 \rangle$ in direct product theory
$\widehat{su}(3)_{k-1} \times \widehat{su}(3)_1 \times \widehat{su}(3)_1$. Tilde indicates   permutation of the   $\widehat{su}(3)_1$ factors.
We will use  the notations $E$, $J$, $\tilde{J}$ for current algebra  generators of above mentioned three factors respectively.

Let us start with identification of    $\phi^{IR}\begin{psmallmatrix}n-1 & n\\n' & n'	\end{psmallmatrix}\phi ^{UV}\begin{psmallmatrix}n & n+1\\n' & n'	\end{psmallmatrix} $ in triple product  theory.
Notice that from (\ref{dim}) and (\ref{WZW_dim}) we have 
\bea
\Delta^{IR}\begin{psmallmatrix}n-1 & n\\n' & n'	\end{psmallmatrix}=
h_{k-1}(\lambda_{n-1,n'})+h_1(\lambda_{2,1})-h_k(\lambda_{n,n'})\,,
\\
\Delta^{UV}\begin{psmallmatrix}n & n+1\\n' & n'	\end{psmallmatrix}=
h_{k}(\lambda_{n,n'})+h_1(\lambda_{2,1})-h_{k+1}(\lambda_{n+1,n'})\,.
\eea
Also using  (\ref{lambda}) one can check that
\bea
\lambda_{n-1,n'}+\lambda_{2,1}+\lambda_{2,1}=\lambda_{n+1,n'}\,.
\eea
Thus there is a single way to construct a 
 highest weight $\lambda_{n+1,n'}$ of total 
 current algebra inside the tensor product
 $V_{n-1,n'}\otimes V_{2,1} \otimes V_{2,1}$. Namely from
 each factor one should  take the corresponding highest weight state
\bea
\label{psi21}
|11 \rangle= |\lambda_{n-1,n'}\rangle|\lambda_{2,1}\rangle |\lambda_{2,1}\rangle\,.
\eea
Notice that $\lambda_{n-1,n'}+\lambda_{2,1}=\lambda_{n,n'}$ so the constraint coming  from the IR coset is also satisfied.

Identification of the remaining states $|i,j\rangle $ are more subtle. The details can be found in App.\ref{app_dom_wall}.
Using the results of this appendix and  (\ref{su3_mod_matrix}), (\ref{mix_coef_dom}), it is easy 
to derive UV/IR mixing coefficients explicitly.
The result for large $k$ is
	\begin{small}
		\bea
		\label{DumWallMixCoef}
			&& \langle \phi^{IR}_1
		\phi_1 | RG \rangle=
		\frac{\sqrt{(n^2-1)((n+n')^2-1 })}{n (n+n')}
		+O\left(\frac{1}{k}\right)\,,
		\\
		\nonumber
		&& \langle \phi^{IR}_1
		\phi_2 | RG \rangle=
		-\frac{1}{n+n'}\sqrt{\frac{(n-1) (n'-1)}{n n'}}
		+O\left(\frac{1}{k}\right)\,,
		\\
		&& \langle  \phi^{IR}_1
		\phi_3| RG \rangle=
		-\frac{1}{n}\sqrt{\frac{(n'+1) (n+n'-1)}{n' (n+n')}}
		+O\left(\frac{1}{k}\right)\,,
		\nonumber
		\\
		&&	\langle \phi^{IR}_2\phi_1| RG \rangle
		=\frac{1}{  (n+ n')}\sqrt{\frac{(n+1) ( n'+1)}{n' n}}+O\left(\frac{1}{k}\right)\,,\nonumber
		\\
		&&	\langle \phi^{IR}_2
		\phi_2 | RG \rangle
		=\frac{ \sqrt{((n+n')^2-1) (n'^2-1)}}{n' (n+n')}+O\left(\frac{1}{k}\right)\,,
		\nonumber
		\\
		&&
		\langle \phi^{IR}_2
		\phi_3| RG \rangle=
		\frac{1}{n'}\sqrt{\frac{(n-1) (n+n'+1)}{n (n+n')}}
		+O\left(\frac{1}{k}\right)\,,
		\nonumber
		\\
		&& \langle \phi^{IR}_3
		\phi_1| RG \rangle=
		\frac{1}{n}\sqrt{\frac{(n'-1) (n+n'+1)}{n' (n+n')}}
		+O\left(\frac{1}{k}\right)\,,
		\nonumber
		\\
		&& \langle \phi^{IR}_3
		\phi_2| RG \rangle=
		-\frac{1}{n'}\sqrt{\frac{(n+1) (n+n'-1)}{n (n+n')}}
		+O\left(\frac{1}{k}\right)\,,
		\nonumber
		\\
		&& \langle \phi^{IR}_3
		\phi_3| RG \rangle=
		\frac{\sqrt{\left(n^2-1\right) \left(n'^2-1\right)}}{n n'}
		+O\left(\frac{1}{k}\right)\,.
		\nonumber
		\eea
	\end{small}
These  is in complete  agreement with our perturbative  result (\ref{set_two_R}). 
This is a strong evidence confirming that  Gaiotto's RG domain wall conjecture is correct also in the higher  rank case.
\section{Conclusions and perspectives}
In this paper we investigated the RG flow between  $A_2^{(p)}$ and
$A_2^{(p-1)}$ minimal models initiated by the relevant field  $\phi \begin{psmallmatrix}	1 & 2  \\	1 & 2	\end{psmallmatrix}$.

We have identified three  classes of RG invariant sets of  local fields (\ref{phi0UV}), (\ref{secund_set_UV}) and  (\ref{uv10}) 
and have shown that in IR limit they flow to the classes  (\ref{phi0IR}),  (\ref{secund_set_IR}) and (\ref{ir10}) respectively.
We obtained  these results 
by computing the  matrices of  anomalies dimensions explicitly using a  well known formula by  A. Zamolodchikov (\ref{anm_dim_mat}). 
Not all structure  constants which enter in this
formula  are available in  literature and their computation was  one
of the main obstacles we had to overcome. Our final results for  matrices of  anomalous dimensions  are
given in 
 (\ref{Gamma0}), (\ref{Gamma2set}) and (\ref{Gamma3set}).

For the second class we were able to drive the IR/UV mixing coefficients (\ref{set_two_R}) by simply finding the eigenvectors 
of  anomalies dimensions matrix $\Gamma$. This result was checked also by constructing the 
RG domain  wall (\ref{DumWallMixCoef}).

Since in the third class there are secondary fields with identical  Virasoro dimensions   the spectrum  of  $\Gamma$ is degenerate.
This is why the linear transformation which diagonalizes $\Gamma$  is not defined uniquely.  Consequently some of UV/IR mixing coefficients
(namely those associated to secondary fields) remain unspecified.
It is reasonable to expect  that this 
uncertainty can be cured by application of  the RG domain wall method.
We hope to address this issue in a future publication.
For a  different  approach to these type of perturbation see  \cite{Sfetsos:2017sep}.

In this paper we introduce the notion of anomalous $W$ weights (see (\ref{Def_WanomDim})) in close analogy with  $\Gamma$. 
The expression  (\ref{W_anm_dim_mat}) for this matrix holds for sets with primary fields only. 
Though our third class includes   secondary fields, nevertheless  we expect that a similar expression should exist in this case too.
 Finding the  mixing coefficients for this class would  be helpful.

While studying the three point functions including  secondary fields we have  encountered an interesting phenomenon: the  violation of  holomorphic factorizability.

Finally it would be interesting to find the UV/IR mixing coefficients for general $W_n$  minimal models. This would allow to investigate 
the large   $k$, $n$ limit with fixed $n/k$ ratio, when in addition a  holographic description is available  \cite{Gaberdiel:2010pz}.
\section*{Acknowledgments}
The  work of H.P. was supported by Armenian SCS grants 21AG‐1C060, 20TTWS-1C035, 20RF-142,
and ANCEF:PS-hepth-2540. R.P acknowledge the support in the framework of Armenian SCS
grants 20RF-142, 21AG‐1C062.
\begin{appendix}
\section{The $\tilde{\phi}$ basis}
\label{ap_phi_lampda}
Consider 	a  basis of primary fields satisfying the condition
 \[\langle \tilde{\phi}_\alpha (1)\tilde{\phi}_\beta(0) \rangle_\lambda=\delta_{\alpha\beta}.\]
 Let us try to find the matrix which connects our initial fields to
this new once. So
\[ \tilde{\phi}_\beta=B_{\beta n}\phi_n \quad {\rm where} \quad  B_{\beta n}=\delta_{\beta n}+\lambda B^{(1)}_{\beta n}\,.\]
	To derive $B^{(1)}$ let us first consider 
	\begin{small}
		\bea
		\label{phi_phi_1}
		&	\langle \phi_\alpha(1)\phi_\beta(0) \rangle_\lambda=
		\langle \phi_\alpha(1)\phi_\beta(0) \rangle-
		\lambda \int \langle \phi_\alpha (1)  \varphi(x) \phi_\beta(0) \rangle d^2x=\\ \nonumber
		&	=\delta_{\alpha \beta}-\lambda \pi C^\alpha_{\,\,\beta}
		\frac{\gamma(\Delta_{\alpha \beta}+\epsilon)\gamma(\Delta_{ \beta \alpha}+\epsilon)}{\gamma(2\epsilon)}=
		\delta_{\alpha \beta}-\lambda \pi C^ \alpha_{\,\,\beta}\frac{2\epsilon}{\epsilon^2-\Delta^2_{\alpha\beta}}\,,
		\eea	
	\end{small}
	where in the last line we have used the fact that $\Delta_{\alpha\beta}$ is  small (of order $\epsilon$).
	Now let us consider
	{\footnotesize
		\bea
		\label{phi_phi_2}
		\langle \phi_\alpha(1)\phi_\beta(0) \rangle_\lambda=
		\langle \tilde{\phi}_\alpha \tilde{\phi}_\beta \rangle-
		\lambda B^{(1)}_{\alpha n}\langle \tilde{\phi}_n \tilde{\phi}_\beta \rangle
		-\lambda B^{(1)}_{\beta n}\langle \tilde{\phi}_\alpha \tilde{\phi}_n \rangle=
		\delta_{\alpha\beta}-\lambda B^{(1)}_{\alpha \beta}-\lambda B^{(1)}_{\beta\alpha}\,.
		\eea
	}
	By compering this to (\ref{phi_phi_1}) we obtain
	\begin{small}
		\bea
		&	B^{(1)}_{\alpha \beta}+ B^{(1)}_{\beta\alpha}=
		\pi C^\alpha_{\,\,  \beta}\frac{2\epsilon}{(\epsilon+\Delta_{\alpha\beta})(\epsilon+\Delta_{\beta\alpha})}=
		\pi C^\alpha_{\,\,  \beta} \left(
		\frac{1}{(\epsilon+\Delta_{\beta\alpha})}+
		\frac{1}{(\epsilon+\Delta_{\alpha\beta})}
		\right)\,.
		\eea 
	\end{small}
	Since $C^\alpha_{\,\, \beta}$ is symmetric the natural solution  is
	$
	B^{(1)}_{\alpha \beta}=
	\frac{\pi C^\alpha_{\,\, \beta}}{(\epsilon+\Delta_{\beta\alpha})}
	$.
	Thus we recover  the result  (\ref{phiTilde}).
\section{Structure constants}
\label{SC}
\subsection{Three point functions}
In this appendix we will be interested in the three point functions 
including secondary fields 
created by the operator  $W_{-1}$.
 The  results of this appendix  will be important for the
  calculation  
 of structure constants of  descendant fields used in the main text. In particular
 we prove here
the result  (\ref{aOffDiagonal}) which is used to calculate the matrix of anomalous $W$-weights.

We adopt standard two point function normalization
\bea
\label{phi2pt}
\langle \phi_1(z_1) \phi_2 (z_2)\rangle= z_{12}^{-2\Delta} \quad {\rm if}\quad 
\Delta_1=\Delta_2\equiv \Delta\,.
\eea
Since in this section we deal with the holomorphic part only for the three point function we will
simply assume that 
\bea
\label{phi3pt}
\langle \phi_1(z_1) \phi_2(z_2) \phi_3(z_3) \rangle= z_{12}^{\Delta_3-\Delta_1-\Delta_2}z_{13}^{\Delta_2-\Delta_1-\Delta_3}
z_{23}^{\Delta_1-\Delta_2-\Delta_3}
\eea
and we will restore actual structure constant $C_{123}$ whenever the ``physical"    correlator, including both holomorphic
and antiholomorphic parts is  considered. Remind the OPE of primary fields with $W$-current
\bea
\label{WphiOPE}
W(z)\phi_i(z_i)=\frac{w_i}{(z-z_i)^3}\phi_i(z_i)+\frac{1}{(z-z_i)^2}W_{-1}\phi_i(z_i)+\frac{1}{z-z_i}W_{-2}\phi_i(z_i)+\dots \,.
\eea
For brevity we will denote the perturbing field as 
\[
\phi_3=\phi \begin{psmallmatrix}1&2\\1&2\end{psmallmatrix}\,.
\]
This field has two independent null vectors at the second level 
	\bea
	\label{xi1}
&&	\left(W_{-2}-\frac{2}{\Delta_3+1}L_{-1}W_{-1}\right)\phi_3=0\,,\\
\label{xi2}
&&	\left(W_{-1}^2+\frac{2 \Delta_3  \left(\Delta_3 +1\right) \left(\Delta_3 +2\right)}{\left(\Delta_3 -3\right) \left(5 \Delta_3 +1\right)}L_{-2}-\frac{3 \left(\Delta_3^2 -1\right) }{2 \left(\Delta_3 -3\right) \left(5 \Delta_3 +1\right)}L_{-1}^2\right)\phi_3=0\,.\qquad
	\eea
Observe that due to  $W_3$ algebra relations
\bea
L_{1}W_{-1}| \phi_3 \rangle =3 w_3 | \phi_3 \rangle =0\,,
\eea
so  that
$W_{-1}| \phi_3\rangle $ is  Virasoro quasi primary.  This is why the three point function takes the form
\bea
\label{Wphi3}
\langle \phi_1 \phi_2 W_{-1}\phi_3 \rangle=a z_{12}^{\Delta_3+1-\Delta_1-\Delta_2}z_{13}^{\Delta_2-\Delta_1-\Delta_3-1}
z_{23}^{\Delta_1-\Delta_2-\Delta_3-1}
\eea
with $a$ being a $z$ independent constant which we will identify below.
Notice that for large $z$ we have
\bea
\label{aspppW}
\langle \phi_1(z_1)\phi_2(z_2)\phi_3(z_3)W(z)\rangle\sim \frac{1}{z^6}\,,
\eea
which guarantees   the vanishing of the  integral
\begin{small}
	\bea
	\label{vanint}
	0=\oint_{{\rm large \, contour}}\langle \phi_1(z_1)\phi_2(z_2)\phi_3(z_3)W(z)\rangle(z-z_1)^2(z-z_2)^2\frac{dz}{2\pi i}=\qquad\qquad\\
	\qquad\qquad=\sum_{k=1}^3\oint_{z_k}\langle \phi_1(z_1)\phi_2(z_2)\phi_3(z_3)W(z)\rangle(z-z_1)^2(z-z_2)^2\frac{dz}{2\pi i}\nonumber\,.
	\eea
\end{small}
Integrals around  the poles can be derived by making use of the OPE (\ref{WphiOPE}). The results are
\begin{small}
	\bea
	\label{arz1}
&&	\oint_{z_1}\langle \phi_1\phi_2\phi_3W(z)\rangle(z-z_1)^2(z-z_2)^2\frac{dz}{2\pi i}=w_1z_{12}^2\langle \phi_1\phi_2\phi_3\rangle\,,
\\
\label{arz2}
&&	\oint_{z_2}\langle \phi_1\phi_2\phi_3W(z)\rangle(z-z_1)^2(z-z_2)^2\frac{dz}{2\pi i}=w_2z_{12}^2\langle \phi_1\phi_2\phi_3\rangle\,,
\\
\label{arz3}
&&	\oint_{z_3}\langle \phi_1\phi_2\phi_3W(z)\rangle(z-z_1)^2(z-z_2)^2\frac{dz}{2\pi i}=\\
&&\qquad	=2 \left(z_{31}z_{32}^2+z_{31}^2z_{32}\right)\langle \phi_1\phi_2W_{-1}\phi_3\rangle+
	z_{31}^2z_{32}^2\langle \phi_1\phi_2W_{-2}\phi_3\rangle\,.
	 \nonumber
	\eea
\end{small}
As it is dictated by (\ref{xi1}) 
\bea
\label{Wmin2_phi}
W_{-2}\phi_3(z_3)=\frac{2}{\Delta_3+1}\frac{\partial}{\partial z_3}W_{-1}\phi_3(z_3)\,.
\eea
Together with (\ref{Wphi3}) we can further simplify  (\ref{arz3})
\bea
\label{intz3}
\oint_{z_3}\langle \phi_1\phi_2\phi_3W(z)\rangle(z-z_1)^2(z-z_2)^2\frac{dz}{2\pi i}=2a \frac{\Delta_{21}}{\Delta_3+1}z_{12}^2 \langle \phi_1\phi_2\phi_3\rangle \,.
\eea
Now we insert (\ref{arz1}), (\ref{arz2}) and (\ref{intz3}) in (\ref{vanint}) to get
\bea
\label{a}
a=\frac{(\Delta_3+1)(w_1+w_2)}{2(\Delta_{1}-\Delta_{2})}\,.
\eea
So for the three point function (\ref{Wphi3}) when $\Delta_1\ne \Delta_2$ we finally get
\bea
\label{Wphi3f}
\langle \phi_1 \phi_2 W_{-1}\phi_3 \rangle=\frac{(\Delta_3+1)(w_1+w_2)}{2(\Delta_{1}-\Delta_{2})}\frac{z_{12}}{z_{13}z_{23}} \langle \phi_1\phi_2\phi_3\rangle\,.
\eea
The case $\Delta_1=\Delta_2$ and $w_1+w_2=0$ should be treated more carefully. In this case the constant $a$ can not be derived through Ward identities. 
We will  address this question   later in this appendix using bootstrap approach.

To derive $\Gamma$  for the third class we need structure constants with descendant fields. In particular we need  to know $\langle (W_{-1} \phi_1)(W_{-1}\phi_2)\phi_3\rangle$. As a first step let us derive  $\langle (W_{-1} \phi_1) \phi_2 \phi_3 \rangle$. 
The technique we use here is the same used for (\ref{Wphi3f}). From (\ref{aspppW}) we observe that
\begin{small}
	\bea
	\label{vanint2}
	0=\oint_{{\rm large \, contour}}\langle W(z) \phi_1\phi_2\phi_3\rangle(z-z_1)(z-z_2)^2(z-z_3)\frac{dz}{2\pi i}=\qquad\\
	\qquad\quad=\sum_{k=1}^3\oint_{z_k}\langle W(z) \phi_1\phi_2\phi_3\rangle(z-z_1)(z-z_2)^2(z-z_3)\frac{dz}{2\pi i}\,.
	\nonumber
	\eea
\end{small}
Inserting the OPE (\ref{WphiOPE}) and evaluating contour integrals  we get
\bea
&&	\oint_{z_1}\langle W(z) \phi_1\phi_2\phi_3\rangle (z-z_1)(z-z_2)^2(z-z_3)\frac{dz}{2\pi i}=\qquad\qquad\\
&&\qquad\qquad\qquad=w_1(z_{12}^2+2z_{12}z_{13})
	\langle \phi_1\phi_2\phi_3\rangle
+z_{12}^2z_{13}	\langle (W_{-1}\phi_1)\phi_2\phi_3\rangle\,,
\nonumber\\
&&	\oint_{z_2}\langle W(z) \phi_1\phi_2\phi_3\rangle (z-z_1)(z-z_2)^2(z-z_3)\frac{dz}{2\pi i}=-
w_2z_{12}z_{23}\langle\phi_1\phi_2\phi_3\rangle\,,
\\
&&	\oint_{z_3}\langle W(z) \phi_1\phi_2\phi_3\rangle (z-z_1)(z-z_2)^2(z-z_3)\frac{dz}{2\pi i}=-
z_{13}z_{23}^2\langle\phi_1\phi_2W_{-1}\phi_3\rangle\,.
\eea
Inserting these expressions  in (\ref{vanint2}) and using   (\ref{Wphi3f})  we obtain
	\bea 
	\label{Wm1p1p2p3}
	&\langle (W_{-1}\phi_1)\phi_2\phi_3\rangle=\left(\left(a+w_2\right)\frac{z_{23}}{z_{12}z_{13}}-w_1\left(\frac{2}{z_{12}}+\frac{1}{z_{13}}\right)\right)\langle \phi_1\phi_2\phi_3\rangle\,.
	\qquad
	\eea
	Similarly
	\bea
	\label{p1Wm1p2p3}
	&\langle \phi_1(W_{-1}\phi_2)\phi_3\rangle=\left(\left(a-w_1\right)\frac{z_{13}}{z_{12}z_{23}}+w_2\left(\frac{2}{z_{12}}-\frac{1}{z_{23}}\right)\right)\langle \phi_1\phi_2\phi_3\rangle\,.
	\eea\
Another correlator we need is
$\langle (W_{-1}\phi_1)\phi_2W_{-1}\phi_3\rangle $. Similar to previous  cases one has
\bea
\label{vanint3}
0=\oint_{{\rm large \, contour}}\langle  W(z)\phi_1\phi_2W_{-1}\phi_3\rangle(z-z_1)(z-z_2)^2(z-z_3)\frac{dz}{2\pi i}=\qquad\\
=\sum_{k=1}^3\oint_{z_k}\langle  W(z)\phi_1\phi_2W_{-1}\phi_3\rangle(z-z_1)(z-z_2)^2(z-z_3)\frac{dz}{2\pi i}\,.
\nonumber
\eea
	On the other hand 
	\bea
	W(z)W_{-1}\phi(0)|0\rangle=\sum_{n\in \mathbb{Z}}\frac{W_n}{z^{n+3}}W_{-1}\phi(0)|0\rangle	=
	\sum_{i=-2}^1\frac{W_i}{z^{i+3}}W_{-1}\phi(0)|0\rangle+O(z)\,.
	\eea
	 Using     $W_3$ algebra
	relations (\ref{comWW}) this OPE can be rewritten as
	\bea
	\label{W(z)Wm1phi}
	&W(z)W_{-1}\phi(0)|0\rangle=\big(\frac{(2-c+32\Delta)\Delta}{(22+5c)}\frac{1}{z^4}
	+\left(wW_{-1}+ \frac{2-c+32\Delta}{22+5c}L_{-1}\right)\frac{1}{z^3}	+ \qquad\\ \nonumber
	&\qquad\qquad\qquad\qquad\qquad\qquad\qquad\qquad+
	W_{-1}^2\frac{1}{z^2}+W_{-2}W_{-1}\frac{1}{z}\big)\phi(0)|0\rangle+O(z)\,.
	\eea
This OPE allows one   to evaluate the contour integral around $z_3$  in (\ref{vanint3}) (integrals around $z_1$ and $z_2$ are standard). Here are the results
\begin{small}
		\bea
&&	\oint_{z_3}\langle  W(z)\phi_1\phi_2W_{-1}\phi_3\rangle(z-z_1)(z-z_2)^2(z-z_3)\frac{dz}{2\pi i}=-z_{13}z_{23}^2\langle\phi_1 \phi_2 W_{-1}^2\phi_3 \rangle+\quad\\
&&\quad+	\left(	-\Delta_3\frac{2-c+32\Delta_3}{22+5c}(2z_{23}+z_{13})+\frac{2-c+32\Delta_3}{22+5c}(2 z_{13}z_{23}+z_{23}^2)\frac{\partial}{\partial z_3}\right)\langle \phi_1 \phi_2 \phi_3\rangle\,,
\nonumber  \\
&& \oint_{z_2}\langle  W(z)\phi_1\phi_2W_{-1}\phi_3\rangle(z-z_1)(z-z_2)^2(z-z_3)\frac{dz}{2\pi i}=-w_2z_{12}z_{23}\langle\phi_1 \phi_2 W_{-1}\phi_3 \rangle\,,
\qquad\quad\\
&& \oint_{z_1}\langle  W(z)\phi_1\phi_2W_{-1}\phi_3\rangle(z-z_1)(z-z_2)^2(z-z_3)\frac{dz}{2\pi i}=\\
&&\qquad\qquad\qquad = w_1(2z_{12}z_{13}+z_{12}^2)\langle\phi_1 \phi_2 W_{-1}\phi_3 \rangle+z_{12}^2z_{13}\langle (W_{-1}\phi_1) \phi_2 W_{-1}\phi_3 \rangle\,.
\nonumber
	\eea
\end{small}
It remains to evaluate $\langle\phi_1 \phi_2 W_{-1}^2\phi_3 \rangle$ using the null vector condition  (\ref{xi2}).
According to conformal Ward identity 
\bea
\langle T(z)\phi_1 \phi_2 \phi_3 \rangle=\sum_{i=1}^{3}
\left(\frac{\Delta_i}{(z-z_i)^2}+\frac{1}{z-z_i}\frac{\partial}{\partial z_i}\right)
\langle\phi_1 \phi_2 \phi_3 \rangle\,.
\eea
Since 
\bea
L_{-2}\phi_3(z_3)= \oint_{z_3}T(z)\phi_3(z_3)(z-z_3)^{-1}\frac{dz}{2\pi i}
\eea
we have
\bea
&\langle \phi_1 \phi_2 L_{-2}\phi_3 \rangle=\sum_{i=1,2}
\left(\frac{\Delta_i}{z_{3i}^2}+\frac{1}{z_{3i}}\frac{\partial}{\partial z_i}\right)
\langle\phi_1 \phi_2 \phi_3 \rangle\,.
\eea
Thus using (\ref{xi2})  we get
{\footnotesize
	\bea
\langle\phi_1 \phi_2 W_{-1}^2\phi_3 \rangle&=&\frac{ \left(\Delta _3+1\right)}{2\left(\Delta _3-3\right) \left(5 \Delta _3+1\right)}
\bigg(3 \left(\Delta _3-1\right) \frac{\partial^2}{\partial z_3^2}-
\\ \qquad 
&-&4 \Delta_3 \left(\Delta _3+2\right)\sum_{i=1,2}\left(\frac{\Delta_i}{z_{3i}^2}+\frac{1}{z_{3i}}\frac{\partial}{\partial z_i}\right)\bigg)\langle\phi_1 \phi_2\phi_3 \rangle.\nonumber\\
\eea
}
Incorporating the above results in (\ref{vanint3}) we obtain 
{\footnotesize
	\bea
&&\langle (W_{-1}\phi_1) \phi_2 W_{-1}\phi_3 \rangle=
\left[
a\left(w_2\frac{z_{23}}{z_{12}z_{13}}-w_1\left(\frac{2}{z_{12}}+\frac{1}{z_{13}}\right)\right)
\frac{z_{12}}{z_{13}z_{23}} \right.
+
\\ \nonumber
&&\quad+	\frac{z_{23}^2 \left(\Delta _3+1\right)}{2z_{12}^{2}\left(\Delta _3-3\right) \left(5 \Delta _3+1\right)}
\left(3 \left(\Delta _3-1\right) \frac{\partial^2}{\partial z_3^2}-4 \Delta_3 \left(\Delta _3+2\right)\sum_{i=1,2}\left(\frac{\Delta_i}{z_{3i}^2}+\frac{1}{z_{3i}}\frac{\partial}{\partial z_i}\right)\right)+\\ \nonumber
&&\quad	\left.+		\Delta_3\frac{2-c+32\Delta_3}{22+5c}\left(2\frac{z_{23}}{z_{12}^2z_{13}}+\frac{1}{z_{12}^2}\right)-\frac{2-c+32\Delta_3}{22+5c}\left(2 \frac{z_{23}}{z_{12}^2}+\frac{z_{23}^2}{z_{12}^2z_{13}}\right)\frac{\partial}{\partial z_3}
\right]\langle\phi_1 \phi_2\phi_3 \rangle\,.
\eea
}
Now we have all ingredients to evaluate  $\langle (W_{-1}\phi_1) (W_{-1}\phi_2) \phi_3 \rangle$. Indeed,  consider the vanishing  integral
\begin{small}
	\bea
	\label{vanint4}
	0=\oint_{{\rm large \, contour}}\langle W(z)(W_{-1} \phi_1)\phi_2W(z)\phi_3\rangle(z-z_1)^2(z-z_2)(z-z_3)\frac{dz}{2\pi i}=\qquad\\
	=\sum_{k=1}^3\oint_{z_k}\langle W(z)(W_{-1} \phi_1)\phi_2W(z)\phi_3\rangle(z-z_1)^2(z-z_2)(z-z_3)\frac{dz}{2\pi i}\,,\nonumber
	\eea
\end{small}
where the summands are equal to
{\footnotesize
\bea
&&\oint_{z_1}\langle W(z)(W_{-1} \phi_1)\phi_2\phi_3\rangle(z-z_1)^2(z-z_2)(z-z_3)\frac{dz}{2\pi i}=\\\nonumber
&&\quad=\frac{2-c+32\Delta_1}{22+5c}\left(\Delta_1(z_{12}+z_{13})+z_{12}z_{13}\frac{\partial}{\partial z_1}\right)\langle\phi_1 \phi_2\phi_3 \rangle+w_1z_{12}z_{13}\langle (W_{-1}\phi_1) \phi_2\phi_3 \rangle\,,\\
&&\oint_{z_2}\langle W(z)(W_{-1} \phi_1)\phi_2\phi_3\rangle(z-z_1)^2(z-z_2)(z-z_3)\frac{dz}{2\pi i}=\\\nonumber
&&\qquad\qquad\qquad=	w_{2}(z_{12}^2-2z_{23}z_{12})\langle (W_{-1} \phi_1)\phi_2\phi_3\rangle+z_{23}z_{12}^2\langle (W_{-1} \phi_1)(W_{-1}\phi_2)\phi_3\rangle\,,\\
&&\oint_{z_3}\langle W(z)(W_{-1} \phi_1)\phi_2\phi_3\rangle(z-z_1)^2(z-z_2)(z-z_3)\frac{dz}{2\pi i}=
-z_{23}z_{13}^2\langle (W_{-1} \phi_1)\phi_2(W_{-1}\phi_3)\rangle\,.\qquad
\eea
}
All three point functions on the right hand sides, besides the one we are interested in, are already in our disposal.
So, from the equality (\ref{vanint4}) we finally get
{\footnotesize
\bea
\label{WfWff}
	\langle (W_{-1} \phi_1)(W_{-1}\phi_2)\phi_3\rangle&=&
\frac{z_{13}^2}{z_{12}^2}\langle (W_{-1} \phi_1)\phi_2(W_{-1}\phi_3)\rangle-
\\
\nonumber 
&-&\left(
w_1\frac{z_{13}}{z_{23}z_{12}}+
w_2\left(\frac{1}{z_{23}}-\frac{2}{z_{12}}\right)
\right)\langle (W_{-1} \phi_1)\phi_2\phi_3\rangle-
\\
\nonumber  
&-&\frac{2-c+32\Delta_1}{(22+5c)z_{23}z_{12}^2}\left(\Delta_1(z_{12}+z_{13})+z_{12}z_{13}\frac{\partial}{\partial z_1}\right)\langle\phi_1 \phi_2\phi_3 \rangle\,.
\eea
}
Actually we are interested in the specific combination
\bea
\nonumber
&\hat{W}\phi:=\left(W_{-1}-\frac{3w}{2\Delta}L_{-1}\right)\phi\,,
\eea
where $w$ and $\Delta$ are the $W$ and Virasoro weights of  $\phi$. From (\ref{Wm1p1p2p3})  we get
\bea
\label{Whfff}
&\langle (\hat{W} \phi^*(z_1))\phi_i(z_2)\phi_3(z_3)\rangle=
\frac{\left(2 a \Delta +\Delta  w+2 \Delta  w_i-3 \Delta _i w+3 \Delta _3 w\right)}{2 \Delta  }
\frac{z_{23}}{z_{12}z_{13}}
\langle\phi^*\phi_i\phi_3\rangle\,.
\eea
Remind  that star stands for conjugate.
We can use  (\ref{WfWff}) together with (\ref{Wm1p1p2p3}) and (\ref{p1Wm1p2p3}) to get
\begin{scriptsize}
	\bea
	\label{WhfWhff}
	\langle(\hat{W} \phi^*)(\hat{W}\phi)\phi_3\rangle&=&
	\bigg(\frac{3 a \left(\Delta _3-1\right) w}{\Delta }-\frac{9  w^2\left(-\Delta _3^2+\Delta _3+2 \Delta \right)}{4 \Delta ^2}-\frac{38  (5 \Delta +1)}{125} -\frac{ \Delta  \left(2 \Delta _3+1\right)}{5}\,
	\\ \nonumber
	& -&\frac{1}{50} \Delta _3 \left(5 \Delta _3-1\right)
	-\frac{(\Delta +3) (5 \Delta +1)}{4 \left(\Delta _3-3\right)}+\frac{9 (5 \Delta +1) (25 \Delta +3)}{500 \left(5 \Delta _3+1\right)}\bigg)
	z_{12}^{-2}\langle \phi^*  \phi \phi_3\rangle\,,
	\eea
\end{scriptsize}
where the relation between  central charge $c$  and  dimension $\Delta_3$
\bea
\nonumber
2-c=\frac{8(1-\Delta_3)^2}{2+\Delta_3}
\eea
is used.
To fix normalization we will need also the two point function of $\hat{W} \phi$, which can be easily derived e.g. considering  
the three point functions (\ref{WfWff}), (\ref{Wm1p1p2p3}) with insertion of the  unit operator instead of  $\phi_3$. Technically  this amounts to replacing $\Delta_3\to 0$ and $a\to 0$. This results in
{\footnotesize
	\bea
	\langle (\hat{W} \phi^*)( \hat{W}\phi) \rangle=\frac{\Delta}{z_{12}^2}\left(\frac{2-c+32\Delta}{22+5c} -\frac{9 w^2}{ 2\Delta^2 }\right)
	\langle \phi^*\phi\rangle\,.
	\eea
}
As one sees from  (\ref{uv10}) we need also three point functions  like 
\begin{small}
	\bea
	\label{3point_function}
&	\langle (\hat{W}\hat{\bar{W}} \phi^*)\phi\phi_3\rangle \,, 
	\quad 
	\langle  \phi^*(\hat{W}\hat{\bar{W}}\phi)\phi_3\rangle \,, 
	\\ \nonumber
&	\langle (\hat{\bar{W}}  \phi^*)(\hat{W} \phi)\phi_3\rangle\,,
	\quad 
	\langle (\hat{W}  \phi^*)(\hat{W} \phi)\phi_3\rangle\,,
	\quad 
	\langle (\hat{W}\hat{\bar{W}}  \phi^*)(\hat{W}\hat{\bar{W}} \phi)\phi_3\rangle\,.
	\eea
\end{small}
The behavior  of a three point function of  these  type  is  somewhat  counter intuitive since
the usual holomorphic anti-holomorphic factorization fails.
From conformal symmetry  
\begin{small}
	\bea
	\label{WWbphi3}
	&\langle \phi^* \phi \hat{\bar{W}}  W_{-1}\phi_3 \rangle=C \langle a \bar{a}  \rangle  |z_{12}|^{2(\Delta_3+1-2\Delta)}|z_{13}z_{23}|^{-2(\Delta_3+1)}\,,
	\eea
\end{small}
where the structure constant of primaries $C$ is explicitly factored out  and the remaining numerical coefficient is denoted as
$\langle a \bar{a}  \rangle$. Actual calculation of this quantity shows that 
\bea
\label{a_and_aa}
\langle a \bar{a}  \rangle \ne a \bar{a}\,.
\eea
From  (\ref{Wm1p1p2p3}) and  (\ref{p1Wm1p2p3}) we obtain 
\bea
& \langle (\hat{W}\hat{\bar{W}} \phi^*)\phi\phi_3\rangle=
\left(4 \langle a \bar{a}  \rangle \Delta ^2+12 a \Delta _3 \Delta  w+9 \Delta _3^2 w^2\right)
\left|\frac{z_{23} }{  z_{12} z_{13}}\right|^2 \langle  \phi^* \phi\phi_3\rangle\,,
\\
& \langle  \phi^*(\hat{W}\hat{\bar{W}} \phi) \phi_3\rangle=
\left(4 \langle a \bar{a}  \rangle \Delta ^2+12 a \Delta _3 \Delta  w+9 \Delta _3^2 w^2\right)
\left|\frac{z_{13} }{  z_{12} z_{23}}\right|^2 \langle  \phi^* \phi\phi_3\rangle\,.
\eea
We can treat the  remaining cases of  (\ref{3point_function}) similarly. 
For example to calculate
\bea
\nonumber
\langle (\hat{W}\hat{\bar{W}}  \phi^*)(\hat{W}\hat{\bar{W}} \phi)\phi_3\rangle
\eea
 one must ``square" (\ref{WhfWhff}), keeping in mind (\ref{a_and_aa}).
 To conclude in this way   we can express the
 structure constants involving descendant fields from the third class (\ref{uv10}) in terms of
 structure constants of primary fields and the constants $a=\bar{a}$ and $\langle a \bar{a}  \rangle$. 
 For brevity let us introduce the notations  
 \[\phi_0:=\phi\begin{psmallmatrix}n & n\\n' & n'\end{psmallmatrix}\]
 and 
 \bea
 \Delta_0=\epsilon ^2 h_0;
 \qquad
h_0= \frac{1}{27}  \left(n^2+n n'+n'^2-3\right)\,,
\\
w_0=\epsilon ^3 s_0\,,
\qquad
s_0=-\frac{ ((n-n') (2 n+n') (n+2 n'))}{243 \sqrt{6}}\,,
\\
\Delta_i=1+ \epsilon h_i, \qquad w_i=\epsilon s_i
\quad {\rm for} \quad 
i=1,2,3,8,9,10\,.
 \eea
For effective structure constants  needed when computing  the matrix
of anomalous dimensions  in third class we have obtained
 \bea
 & C^4_{\,\, i}=C^0_{\,\, i} \frac{\left(h_i+1\right)^2}{2 h_0}\,,
 \\
 & C^6_{\,\, i}= C^5_{\,\, i}=C^0_{\,\, i} 
 \frac{(h_i+1) (3 s_0 (h_i+1)-s_i h_0 (2 h_i+1))}{h_0 \sqrt{\frac{2}{3} h_0^2 (12 h_0+1)-36 s_0^2}}\,,
 \\
 & C^7_{\,\, i}=C^0_{\,\, i} \frac{3 (s_i h_0 (2 h_i+1)-3 s_0 (h_i+1))^2}{-54 s_0^2 h_0+12 h_0^4+h_0^3}\,,
 \eea
  where the explicit form of  $h_i$ is determined from  (\ref{set_3_dim}).
 For the  structure constants $C^i_j$ with $i,j\in \{4,5,6,7\}$ we  have used the explicit expressions (to be derived later on): 
 \bea
 \label{a_coef}
& a= -\frac{\epsilon  ((n'-n) (2 n+n') (n+2 n'))}{6 \sqrt{6} \left(n^2+n n'+n'^2-3\right)}+O(\epsilon^2),
 \\
\label{aa_coef}
& \langle a \bar{a}  \rangle =\frac{\epsilon ^2 \left(4 \left(n^2+n n'+n'^2-3\right)+9\right)}{27\ 8}+O(\epsilon^3)\,.
 \eea
 To avoid confusion let  us emphasize that this result is valid only in current case i.e. when both fields are descendants of $\phi_0$.
 Here are the final expressions
 \bea
& C^4_{\,\, 4}=C^0_{\,\, 0}\left(\frac{729 \epsilon ^{-2}}{4  \left(n^2+n n'+n'^2-3\right)^2}+O\left(\frac{1}{\epsilon}\right)\right)\,,
 \\
& C^4_{\,\, 5}=C^4_{\,\, 6}=C^0_{\,\, 0}\left(1+O(\epsilon^{-1})\right)\,,
 \\
& C^4_{\,\, 7}=C^4_{\,\, 4}+O(\epsilon)\,,
 \eea
 {\footnotesize
 \bea
& C^5_{\,\, 5}= C^5_{\,\,6}= C^6_{\,\, 6}= C^4_{\,\, 4}+O(\epsilon)\,,
 \\
& C^5_{\,\,7}= C^6_{\,\, 7}  =-C^0_{\,\, 0}\,,
\left(
\frac{81 \sqrt{3} \epsilon ^{-2}(n-n') (2 n+n') (n+2 n')}{2  \sqrt{\left(n^2-1\right) \left(n'^2-1\right) \left((n+n')^2-1\right)} \left(n^2+n n'+n'^2-3\right)^2}+O\left(\frac{1}{\epsilon}\right)
\right)\,,
 \eea
}
 {\footnotesize
 \bea
& C^7_{\,\, 7}=
 C^0_{\,\, 0}
 \left(\frac{27 \left(4 \left(n^2+n n'+n'^2-3\right)^3+9 \left(n^2+n n'+n'^2-3\right)^2+3 (n-n')^2 (2 n+n')^2 (n+2 n')^2\right)}{4 \epsilon ^2 \left(n^2-1\right) \left(n'^2-1\right)  \left((n+n')^2-1\right) \left(n^2+n n'+n'^2-3\right)^2}+O\left(\frac{1}{\epsilon}\right)\right)\,.
 \eea
}
\subsection{OPE  up to level one}
In order to identify  $a$ and $\langle a \bar{a}  \rangle$ from four point correlation function
we need the  next to leading order terms of  OPE expansion explicitly.
It follows from OPE 
\bea
T(z)\phi (0)=\frac{\Delta}{z^2}\phi(0)+\frac{1}{z}(\partial\phi)(0)+...
\eea
and (\ref{WphiOPE}) \cite{Belavin:1984vu,Zamolodchikov:1985wn} that
\bea
&&\left[L_n,\phi(z) \right]=z^n\left((n+1)\Delta+z\partial_z\right)\phi(z)\,,\\
\label{comWnPhi}
&&\left[W_n,\phi(z) \right]=z^n\left(\frac{(n+2)(n+1)}{2}w+(n+2)zW_{-1}+z^2W_{-2}\right)\phi(z)\,.
\eea
For brevity let us denote 
\[\phi_{-b \omega_1}= \phi \begin{psmallmatrix}1 & 2\\1 & 1	\end{psmallmatrix}\equiv \phi_1\,.\]
This is a doubly degenerated primary  field with independent null vectors 
\bea
\label{x1}
&&\left(W_{-1}-\frac{3w_1}{2\Delta_1}L_{-1}\right)\phi_{1}=0\,,
\\
\label{x2}
&&\left(W_{-2}-\frac{12w_1}{\Delta_1(5\Delta_1+1)}L_{-1}^2+
\frac{6w_1(\Delta_1+1)}{\Delta_1(5\Delta_1+1)}L_{-2}\right)\phi_{1}=0\,.
\eea
From general arguments one has\footnote{ We have suppressed the contribution of unit operator. For our proposes it is sufficient to concentrate on $W$-block $[\varphi]$ only.}
\bea
\label{OPEptildep}
\phi^*_{1}(z)\phi_{1}(0)=z^{\Delta-2\Delta_1}\left(\varphi(0)+z\left(\alpha L_{-1}+\beta W_{-1}\right)\varphi(0)+...\right)\,,
\eea 
where as usual  $\varphi\equiv  \phi 
\begin{psmallmatrix}
1 & 2\\
1 & 2	
\end{psmallmatrix}$. We can act on  This OPE with $L_1$ and find that $\alpha=1/2$. 
Similarly to find $\beta$ we  act on  (\ref{OPEptildep}) with $W_1$ 
\bea
\label{W1OPEptildep}
\left[W_1,\phi^*_{1}(z)\right]\phi_{1}(0)&=&\beta z^{\Delta-2\Delta_1+1}\left[W_1,W_{-1}\right]\varphi(0)
=
\\ \nonumber
&=&\beta z^{\Delta-2\Delta_1+1}
\left(-\frac{1}{5}\Delta+\frac{32}{22+5c}\left(\frac{1}{5}\Delta+\Delta^2\right)\right)\varphi(0)\,,
\eea
where in  second row  the commutation relation  (\ref{comWW}) is used.
On the other hand
using  (\ref{comWnPhi}) and (\ref{x1})  for the lhs of (\ref{W1OPEptildep}) we have
\begin{small}
	\bea
	\label{W1_phi1_com2}
	&\left[W_1,\phi^*_{1}(z)\right]\phi_{1}(0)=-\left(3w_1z+z^2 \frac{9w_1}{2\Delta_1}\partial_z\right)\phi^*_{1}(z)\phi_{1}(0)
	+z^3 (W_{-2}\phi^*_{1}(z))\phi_{1}(0)\,.
	\eea
\end{small}
From null vector condition (\ref{x2})
\begin{small}
	\bea
	\label{Wm2PP}
&	(W_{-2}\phi^*_{1}(z))\phi_{1}(0)=-\frac{12w_1}{\Delta_1(5\Delta_1+1)}\partial_z^2
	\phi^*_{1}(z)\phi_{1}(0)+\frac{6w_1(\Delta_1+1)}{\Delta_1(5\Delta_1+1)} (L_{-2}\phi^*_{1}(z))\phi_{1}(0)\,.
	\eea
\end{small}
The second summand of this expression  can be refurnished  as
\begin{small}
	\bea
	\label{Lm2ftf}
&(L_{-2}\phi^*_{1})(z)\phi_{1}(0)=	\oint_{z}T(\zeta)(\zeta-z)^{-1}\phi^*_{1}(z)\phi_{1}(0)\frac{d\zeta}{2\pi i}=\\
&	\oint_{z,0}z^{\Delta-2\Delta_1}(\zeta-z)^{-1}T(\zeta)\varphi(0)\frac{d\zeta}{2\pi i}-
	\oint_{0}(\zeta-z)^{-1}\phi^*_{1}(z)\left(\frac{\Delta_1}{\zeta^2}\phi_{1}(0)+\frac{1}{\zeta}(\partial\phi_{1})(0)\right)\frac{d\zeta}{2\pi i}\,,
	\nonumber
	\eea
\end{small}
where besides the definition of $L_{-2}$ in first term of the second row    OPE (\ref{OPEptildep}) is applied.
Using $T\varphi$ OPE and performing $\zeta$ integration around $0$ and $z$ one can easily get convinced that it is of order
$z^{\Delta-2\Delta_1}$. This term can be ignored since we are interested in more singular terms of order $z^{\Delta-2\Delta_1-2}$  (notice the factor $z^3$ in front of second term in (\ref{W1_phi1_com2})). The remaining integral around zero in (\ref{Lm2ftf}) is of desired order:
\bea
\label{int_z0}
&&-
\oint_{0}(\zeta-z)^{-1}\phi^*_{1}(z)\left(\frac{\Delta_1}{\zeta^2}\phi_{1}(0)+\frac{1}{\zeta}(\partial \phi_{1})(0)\right)\frac{d\zeta}{2\pi i}=
\\ \nonumber
&& \qquad
\frac{\Delta_1}{z^2}\phi^*_{1}(z)\phi_{1}(0)+\frac{1}{z}\phi^*_{1}(z)\partial \phi_{1}(0)
=(3\Delta_1-\Delta) z^{\Delta-2\Delta_1-2} \varphi(0)\,.
\eea
Combining (\ref{int_z0}), (\ref{Lm2ftf}) and (\ref{Wm2PP}) we will get
{\footnotesize
\bea
&(W_{-2}\phi^*_{1}(z))\phi_{1}(0)=\frac{-12w_1(\Delta-2\Delta_1)(\Delta-2\Delta_1-1)+6w_1(\Delta_1+1)(3\Delta_1-\Delta)}{\Delta_1(5\Delta_1+1)}
z^{\Delta-2\Delta_1-2} \varphi(0)+...\,.
\eea
}
Now we can  find  $\beta$ by  inserting this in (\ref{W1_phi1_com2}) and compering the result with (\ref{W1OPEptildep})
\bea
\beta=-\frac{3 (5 c+22) \left(13 \Delta_1 -8 \Delta+1\right) }{2 \Delta_1  (5 \Delta_1 +1) \left(c-32 \Delta-2\right)}w_1\,.
\eea
Taking into account that  
\begin{small}
	\bea
	\Delta_1 = \frac{p-3}{3 (p+1)}\,,\qquad
	w_1= \frac{(p-3) }{9 (p+1)}\sqrt{\frac{2 (2 p-3)}{3 (2 p+5)}}
	\eea
\end{small}
and expressing $c$ and $\Delta$ in terms of $p$ we finally  get
\bea
\label{beta}
\beta=-\frac{\sqrt{(2 p-3)(2 p+5)}}{\sqrt{6} (2 p-1)}\,.
\eea
\subsection{Structure constants from 4-point functions}
\label{AppStCons}
In this section we closely follow \cite{Fateev:2007ab}, where $A_2$ Toda field theory 
is thoroughly analyzed.  Whenever necessary
we make appropriate modifications having in mind application for 
minimal models. 

Consider the correlation function
\bea
\label{corr}
\langle 
\phi_{\alpha_2}(\infty)\phi_{\kappa \omega_2}(1)\phi_{-b \omega_1}(x)\phi_{\alpha_1}(0)
\rangle 
=(x{\bar x})^{b(\alpha_1\cdot h_1)}((1-x)(1-{\bar x}))^{\frac{b\kappa}{3}}G(x,{\bar x})\,.
\eea
The function $G$ can be expressed in terms of the generalized hypergeometric function 
${}_3F_2$ as follows:
\bea
\label{4point_s} 
G(x,{\bar x})=\sum_{i=1}^3s_i|F^{(s)}_i(x)|^2\, ,
\eea 
where
\begin{small}
\bea 
\label{Fs_1}
&&F^{(s)}_1(x)=\pFq[6]{3}{2}{A_1;,A_2;,A_3}{B_1;,B_2}{x}\,,
\\
\label{Fs_2}
&&F^{(s)}_2(x)=x^{1-B_1}\pFq[6]{3}{2}{1-B_1+A_1;,1-B_1+A_2;,1-B_1+A_3}{2-B_1;,1-B_1+B_2}{x}\,,
\\
\label{Fs_3}
&&F^{(s)}_3(x)=x^{1-B_2}\pFq[6]{3}{2}{1-B_2+A_1;,1-B_2+A_2;,1-B_2+A_3}{2-B_2;,1-B_2+B_1}{x}\,.
\eea
The parameters are given by ($i=1,2,3$):
\bea
\label{AB_parametrization} 
&&A_i=\frac{b (\kappa-2 b)}{3}+b \left(\alpha _2-Q\right)\cdot h_i+b \left(\alpha _1-Q\right)\cdot h_1\,,
\nonumber\\
&&B_1=1+b \left(\alpha _1-Q\right)\cdot\left(h_1-h_2\right)\,,
\nonumber\\
&&B_2=1+b \left(\alpha _1-Q\right)\cdot\left(h_1-h_3\right)\,.
\eea 
\end{small} 
We'll call eq. (\ref{4point_s}) the s-channel representation of the correlation function 
since $x\sim 0$ behavior (in particular the single valuednes) is quite transparent.
The CFT blocks $F^{(s)}_i$, for $i=1,2,3$ correspond to intermediate primaries 
$\phi_{\alpha_1-b\omega_1}$, $\phi_{\alpha_1+b\omega_1-b\omega_2}$ and $\phi_{\alpha_1+b\omega_2}$
respectively, as can be easily seen by examining dimensions.

 Using 
the transformation formula under $x\to1/x$
\begin{small}
	\bea
	\label{Fsu}
	\frac{\Gamma (a) \Gamma (b) \Gamma (c)}
	{\Gamma (e) \Gamma (f)}&&\pFq[2]{3}{2}{a;,b;,c}{e;,f}{x}=\\
	&=&\frac{\Gamma (a) \Gamma (b-a) \Gamma (c-a)}{\Gamma (e-a) \Gamma (f-a)}\,\,
	x^{-a}\,\pFq[2]{3}{2}{a;,1-e+a;,1-f+a}{1-b+a;,1-c+a}{\frac{1}{x}}\nonumber\\
	&+&\,\frac{\Gamma (b) \Gamma (a-b) \Gamma (c-b)}{\Gamma (e-b) \Gamma (f-b)}\,\,
	x^{-b}\,\pFq[2]{3}{2}{b;,1-e+b;,1-f+b}{1-a+b;,1-c+b}{\frac{1}{x}}\nonumber\\
	&+&\,\frac{\Gamma (c) \Gamma (a-c) \Gamma (b-c)}{\Gamma (e-c) \Gamma (f-c)}\,\,
	x^{-c}\,\pFq[2]{3}{2}{c;,1-e+c;,1-f+c}{1-a+c;,1-b+c}{\frac{1}{x}}\qquad
	\nonumber
	\eea
\end{small}
eq. (\ref{4point_s}) alternatively can be represented as (u-channel representation) 
\bea
\label{4point_u} 
G(x,{\bar x})=\sum_{i=1}^3u_i|F^{(u)}_i(x)|^2\, ,
\eea 
where
\begin{small}
\bea 
\label{Fu_1}
&&F^{(u)}_1(x)=x^{-A_1}\pFq[6]{3}{2}{A_1;,1-B_1+A_1;,1-B_2+A_1}{1-A_2+A_1;,1-A_3+A_1}{\frac{1}{x}}\,,
\\
\label{Fu_2}
&&F^{(u)}_2(x)=x^{-A_2}\pFq[6]{3}{2}{A_2;,1-B_1+A_2;,1-B_2+A_2}{1-A_1+A_2;,1-A_3+A_2}{\frac{1}{x}}\,,
\\
\label{Fu_3}
&&F^{(u)}_3(x)=x^{-A_3}\pFq[6]{3}{2}{A_3;,1-B_1+A_3;,1-B_2+A_3}{1-A_1+A_3;,1-A_2+A_3}{\frac{1}{x}}\,.
\eea
\end{small} 
Consideration similar to the s-channel case ensures that the CFT blocks $F^{(u)}_i$, for $i=1,2,3$ 
correspond to u-channel intermediate primaries 
$\phi_{\alpha_2-b\omega_1}$, $\phi_{\alpha_2+b\omega_1-b\omega_2}$ and $\phi_{\alpha_2+b\omega_2}$.
The absence of cross terms in (\ref{4point_u}) is guaranteed provided 
$s_1$, $s_2$, $s_3$ are related as \cite{Fateev:2007ab}
\bea 
\label{sratios}
s_2:s_1=\frac{\gamma (B_1)\gamma (B_2)}{\gamma (2-B_1)\gamma (1-B_1+B_2)}\prod_{i=1}^3
\frac{\gamma (1-B_1+A_i)}{\gamma (A_i)}\,,
\nonumber\\
s_3:s_1=\frac{\gamma (B_1)\gamma (B_2)}{\gamma (2-B_2)\gamma (1-B_2+B_1)}\prod_{i=1}^3
\frac{\gamma (1-B_2+A_i)}{\gamma (A_i)}\,.
\eea 
Thus such choice ensures single valuedness of correlation function around  
$x\sim \infty$ too. The single valuedness around remaining singularity at $x\sim 1$ now is
automatically guaranteed, since a small cycle surrounding $x=1$ is equivalent to the
difference of cycles around $x=\infty$ and $x=0$. Using (\ref{Fsu}) one can show that 
the coefficients $u_i$ are related to $s_1$:
\bea 
\label{uratios}
u_1:s_1=\frac{\gamma \left(A_2-A_1\right) \gamma \left(A_3-A_1\right) \gamma \left(B_1\right) 
\gamma \left(B_2\right)}{\gamma \left(A_2\right) \gamma \left(A_3\right) 
\gamma \left(B_1-A_1\right) \gamma \left(B_2-A_1\right)}\,,
\nonumber\\
u_2:s_1=\frac{\gamma \left(A_1-A_2\right) \gamma \left(A_3-A_2\right) \gamma \left(B_1\right) 
\gamma \left(B_2\right)}{\gamma \left(A_1\right) \gamma \left(A_3\right) 
\gamma \left(B_1-A_2\right) \gamma \left(B_2-A_2\right)}\,,
\\
u_3:s_1=\frac{\gamma \left(A_1-A_3\right) \gamma \left(A_2-A_3\right) \gamma \left(B_1\right) 
\gamma \left(B_2\right)}{\gamma \left(A_1\right) \gamma \left(A_2\right) 
\gamma \left(B_1-A_3\right) \gamma \left(B_2-A_3\right)}\,.
\nonumber
\eea 
It is not surprising that the correlation function is fixed up to an overall constant 
factor, since we have note imposed any field normalization condition yet.
  
To calculate the OPE structure constants of our interest, let us start with the choice 
\bea 
\label{restriction_kappa_alpha_2}
\kappa=-b; \qquad \alpha_2=\alpha^*_1\,.
\eea For later use let us emphasize that in this 
case the parameters of  hypergeometric functions become related as
\bea
\label{ABrelation} 
B_1=1+A_2-A_3\,,\qquad B_2=1+A_1-A_3.
\eea 
The t-channel (i.e. $x\sim 1$) in this case gets contribution 
of the unit operator which emerges in OPE 
\begin{small}
	\bea 
	\label{OPEtch}
	\phi_{-b\omega_1}(z)\phi_{-b\omega_2}(1)\sim |z-1|^{-4\Delta(-b\omega_1)}[I]+
	|z-1|^{2\Delta(-be_0)-4\Delta(-b\omega_1)}C_{-b\omega_1,-b\omega_2}^{-be_0}[\phi_{-be_0}].\qquad
	\eea
\end{small}
Then imposing canonical unit-normalization condition on 
two-point functions we can remove the remaining ambiguity thus completely fixing the 4-point 
correlation function. 
Behavior of the hypergeometric function around 
\[x\sim 1, \qquad \arg (1-x)<\pi \]
is given by the formula
\bea 
\label{xsim1} 
\frac{\Gamma(a)\Gamma(b)\Gamma(c)}{\Gamma(e)\Gamma(f)}\pFq[4]{3}{2}{a,b,c}{e,f}{x}=
\sum_{n=0}^\infty g_n(0)(1-x)^n+\sum_{n=0}^\infty g_n(s)(1-x)^{n+s}\,,
\eea 
where $s=e+f-a-b-c$. The general formulae for the coefficients could be found in  \cite{Bhring:1992}.
For our purposes we need only the coefficients $g_0(0)$, $g_1(0)$ and $g_0(s)$ (using 
obvious symmetry we have exchanged the roles of $b$ and $c$ compared to \cite{Bhring:1992} for 
later convenience):
\bea 
&&g_0(s)=\Gamma(-s)\,,
\nonumber\\
&&g_0(0)=\frac{\Gamma (a) \Gamma (c) \Gamma (s)}{\Gamma (a+s) \Gamma (c+s)}
\pFq[2]{3}{2}{e-b;,f-b;,s}{a+s;,c+s}{1}\,,
\\
&&g_1(0)=\frac{\Gamma (a+1) \Gamma (c+1) \Gamma (s-1)}{\Gamma (a+s) \Gamma (c+s)}
\pFq[2]{3}{2}{e-b;,f-b;,s-1}{a+s;,c+s}{1}\,.
\nonumber
\eea 
The hypergeometric function of argument $1$ should be understood as the result 
of analytic continuation from a region of parameters, where the hypergeometric series 
converges. It is well known, that
\bea 
\label{Def_hypF}
\pFq[2]{3}{2}{a;,b;,c}{e;,f}{x}=\sum_{n=0}^\infty \frac{(a)_n(b)_n(c)_n}{ (e)_n(f)_n n!}\,x^n
\eea  
converges at $x=1$ provided $e+f-a-b-c>0$. If we apply  (\ref{xsim1}) to investigate $x\sim 1$ 
limit of the correlation function (\ref{4point_s}), one generically gets 
hypergeometric functions of unit argument which do not meet the convergence condition. 
Fortunately this difficulty can be overcome  performing one more transformation
using the identity
\bea 
\pFq[2]{3}{2}{a;,b;,c}{e;,f}{1}=\frac{\Gamma (f) \Gamma (-a-b-c+e+f) 
}{\Gamma (f-a) \Gamma (-b-c+e+f)}\,\pFq[2]{3}{2}{a;,e-b;,e-c}{e;,a+s}{1}
\eea 
for the  cases (\ref{Fs_1}), (\ref{Fs_2}) and the identity (obtained from the  previous one by reshuffling $e$ with $f$)  
\bea 
\pFq[2]{3}{2}{a;,b;,c}{e;,f}{1}=\frac{\Gamma (e) \Gamma (-a-b-c+e+f) 
}{\Gamma (e-a) \Gamma (-b-c+e+f)}\,\pFq[2]{3}{2}{a;,f-b;,f-c}{f;,a+s}{1}
\eea
for the case (\ref{Fs_2}). These formulae can be derived combining Lemma 1 of 
\cite{Bhring:1992} with obvious symmetry properties of the hypergeometric function 
with respect to its parameters. Interestingly in all three cases 
one arrives at the unit argument hypergeometric functions with 
$s=1+b^2= 1/(p+1) >0$, so that convergence criteria is fulfilled.

For the moment let us concentrate on the second simpler summand of expansion (\ref{xsim1}), 
which behaves as $(1-x)^{-s}$ times something analytic at $x\sim 1$. 
The main contribution of this part in (\ref{4point_s}) is 
equal to (remind that $B_{1,2}$ are related to $A_{1,2,3}$ via (\ref{ABrelation}) )
{\footnotesize
\bea
\label{unitcontamplitude}
C&=&\Gamma(3\rho -2)^2\\
&\times&\left(\frac{s_1\Gamma^2(B_1)\Gamma^2(B_2)}{\prod_{i=1}^3\Gamma^2(A_i)}
+\frac{s_2\Gamma^2(2-B_1)\Gamma^2(1-B_1+B_2)}{\prod_{i=1}^3\Gamma^2(1-B_1+A_i)}
+\frac{s_3\Gamma^2(2-B_2)\Gamma^2(1-B_2+B_1)}{\prod_{i=1}^3\Gamma^2(1-B_2+A_i)}\right).\nonumber
\eea 
}
Notice also that the exponent of prefactor   
$
|1-x|^{\frac{b\kappa}{3}}\,,\nonumber  
$
relating the correlator (\ref{corr}) to (\ref{4point_s}), in our case is equal to 
$ 
\frac{b\kappa}{3}=-\frac{b^2}{3}=-4\Delta (-b\omega_1)
$, which exactly is the exponent expected for the contribution of unit operator 
(see (\ref{OPEptildep})). Under this circumstances The 4-point correlator factorizes 
as
\bea 
\langle 
\phi_{\alpha^*_1}(\infty)\phi_{-b\omega_2}(1)\phi_{-b \omega_1}(x)\phi_{\alpha_1}(0)
\rangle\sim 
\langle \phi_{\alpha^*_1}(\infty) \phi_{\alpha_1}(0)\rangle 
\langle \phi_{-b\omega_1}(x) \phi_{-b\omega_2}(1)\rangle \,.
\eea
Thus standard unit normalization for 2-point functions is compatible with  
(\ref{unitcontamplitude}) if $C\equiv 1$.   
Combining this condition with (\ref{sratios}) we fix
\bea 
&&s_1=\frac{\gamma \left(A_1\right) \gamma \left(A_2\right) 
\gamma \left(3-3 A_3\right) \gamma \left(A_3\right)}
{\gamma \left(A_1-A_3+1\right) \gamma \left(A_2-A_3+1\right)}\,,
\\
&&s_2=\frac{\gamma \left(2 A_3-A_2\right) \gamma \left(A_1+A_3-A_2\right) 
\gamma \left(A_3\right) \gamma \left(3-3 A_3\right)}
{\gamma \left(A_1-A_2+1\right) \gamma \left(A_3-A_2+1\right)}\,,
\\
&&s_3=\frac{\gamma \left(2 A_3-A_1\right) \gamma \left(A_2+A_3-A_1\right) 
\gamma \left(A_3\right) \gamma \left(3-3 A_3\right)}
{\gamma \left(A_2-A_1+1\right) \gamma \left(A_3-A_1+1\right)}\,.
\eea 
Using (\ref{uratios}) we can easily get also the u-channel constants
\bea 
&&u_1=\frac{\gamma \left(A_1-A_2+A_3\right) \gamma \left(A_1\right) 
\gamma \left(A_3\right) \gamma \left(3-3 A_3\right)}
{\gamma \left(A_1-A_2+1\right)\gamma \left(A_1-A_3+1\right)}\,,
\\
&&u_2=\frac{\gamma \left(A_2-A_1+A_3\right) \gamma \left(A_2\right) 
\gamma \left(A_3\right) \gamma \left(3-3 A_3\right)}
{\gamma \left(A_2-A_1+1\right) \gamma \left(A_2-A_3+1\right)}\,,
\\
&&u_3=\frac{\gamma \left(2 A_3-A_1\right) \gamma \left(2 A_3-A_2\right) 
\gamma \left(A_3\right) \gamma \left(3-3 A_3\right)}
{\gamma \left(A_3-A_1+1\right) \gamma \left(A_3-A_2+1\right)}\,.
\eea
Remind that the s- and t- channel intermediate fields are already identified 
hence the coefficients $s_i$, $b_i$ are just the (squared) OPE structure constants
\bea 
\label{str_s}
s_1=\left[C_{-b\omega_1,\alpha_1}^{\alpha_1-b\omega_1}\right]^2\,,\quad 
s_2=\left[C_{-b\omega_1,\alpha_1}^{\alpha_1+b\omega_1-b\omega_2}\right]^2\,,\quad 
s_3=\left[C_{-b\omega_1,\alpha_1}^{\alpha_1+b\omega_2}\right]^2\,
\eea
and
\bea
\label{str_u} 
u_1=\left[C_{-b\omega_1,\alpha_2}^{\alpha_2-b\omega_1}\right]^2\,,\quad 
u_2=\left[C_{-b\omega_1,\alpha_2}^{\alpha_2+b\omega_1-b\omega_2}\right]^2\,,\quad 
u_3=\left[C_{-b\omega_1,\alpha_2}^{\alpha_2+b\omega_2}\right]^2\,.
\eea 
Using (\ref{AB_parametrization}) and (\ref{ABrelation}) one can check that both 
(\ref{str_s}) and (\ref{str_u}) consistently lead to the formulae
(\ref{str_simple1}), (\ref{str_simple2}) and (\ref{str_simple3}). 

To find structure constants including the perturbing field $\phi_{-b(\omega_1+\omega_2)}$, 
it is necessary to investigate the t-channel contributions coming from the first summand 
of the formula (\ref{xsim1}). Besides the leading terms we shall keep also the 
subleading contributions, since we are also interested in the structure constants of 
certain first level descendant fields.
\begin{scriptsize}
\bea 
\label{C_diagonal}
&&C_{-b\omega_1,-b\omega_2}^{-b(\omega_1+\omega_2)}
C_{-b(\omega_1+\omega_2),\alpha_1,\alpha^*_1}=
\frac{\sin\left(\pi A_3\right)\gamma^2(A_3) 
\gamma \left(2-3 A_3\right) \gamma\left(3-3 A_3\right)}{\sin\left(3\pi A_3\right)}\\
&&\times\left(\frac{\sin(\pi (A_1-3 A_3))\gamma(A_1)\gamma(A_2)F_1^2}
{\sin(\pi(A_1-A_3)) \gamma (A_1-3 A_3+2) \gamma(A_2-A_3+1)}
+\frac{\sin(\pi(A_1-A_2-2A_3))\gamma(A_2-A_3)\gamma(A_1-A_2+A_3)F_2^2}
{\sin(\pi(A_1-A_2))\gamma(A_2-2A_3+1)\gamma(A_1-A_2-2 A_3+2)}\right.\nonumber\\
&&\hspace{7cm}+\left.\frac{\sin(\pi(A_1+A_3))\gamma(A_1-A_2) 
\gamma(A_2+A_3-A_1)F_3^2}{\sin(\pi(A_1-A_3))\gamma(A_1-2 A_3+1)
\gamma(2-A_1-A_3)}\right)\,,\nonumber
\eea
\end{scriptsize} 
where
\begin{small}
\bea \nonumber
&&F_1=\pFq[2]{3}{2}{A_1;,1+A_2-2A_3;,1-A_3}{2+A_1-3A_3;,1+A_2-A_3}{1}\,,
\\
&&F_2=\pFq[2]{3}{2}{1-A_2;,1-A_3;,A_1-A_2+A_3}{2+A_1-A_2-2A_3;,1-A_2+A_3}{1}\,,
\\
&&F_3=\pFq[2]{3}{2}{1-A_3;,1-A_1+A_2-A_3;,-A_1+2A_3}{1-A_1+A_2;,2-A_1-A_3}{1}\,.
\nonumber
\eea
\end{small} 
The structure constant $C_{-b\omega_1,-b\omega_2}^{-b(\omega_1+\omega_2)}$ is easily 
calculated using the special case $\alpha_1=-b\omega_2$ of the first equality in (\ref{str_s}):
\bea 
C_{-b\omega_1,-b\omega_2}^{-b(\omega_1+\omega_2)}=
\sqrt{\frac{\gamma(2-3\rho)\gamma(3-3\rho )}{\gamma(3-4\rho)\gamma(2-2\rho)}}\,.
\eea 
Thus, in principle, (\ref{C_diagonal}) determines the diagonal structure constants 
of type 
\[C_{-b(\omega_1+\omega_2),\alpha,\alpha^*}\,.\]
 Unfortunately, in general case 
the generalized hypergeometric functions with unit arguments in (\ref{Cdiag}) can not 
be expressed via gamma-functions and we do not know if there is a way to simplify further these
expressions. Fortunately in large $p$ limit, which is considered in this paper the (rather subtle) calculations lead to concise algebraic expressions. 
Notice also, that in an important case $\alpha=-b(\omega_1+\omega_2)$, above mentioned 
generalized hypergeometric functions of unit argument can be represented in terms of 
gamma-functions using so called Watson's  sum:
{\footnotesize
\bea 
\pFq[2]{3}{2}{a;,b;,c}{\frac{1}{2}(a+b+1);,2c}{1}
=\frac{\Gamma\left(\frac{1}{2}\right)\Gamma\left(\frac{1}{2}+c\right)
	\Gamma\left(\frac{1}{2}(a+b+1)\right)\Gamma\left(c+\frac{1}{2}(1-a-b)\right)}
{\Gamma\left(\frac{1}{2}(a+1)\right)\Gamma\left(\frac{1}{2}(b+1)\right)
	\Gamma\left(c+\frac{1}{2}(1-a)\right)\Gamma\left(c+\frac{1}{2}(1-b)\right)}\,.\qquad
\eea
}
Calculation is rather lengthy, but the final expression has a nice factorized form
\begin{small}
	\bea 
	C_{-b(\omega_1+\omega_2),-b(\omega_1+\omega_2)}^{-b(\omega_1+\omega_2)}=
	\frac{2 (4-5 \rho )^2\gamma \left(\frac{\rho }{2}\right) 
		\gamma \left(\frac{5 \rho }{2}-2\right) \sqrt{\gamma (4-4 \rho ) \gamma (2-2 \rho )}}
	{(3 \rho -2) (4 \rho -3) \gamma (3-3 \rho ) \gamma \left(\frac{3 \rho }{2}-1\right)^2}\,.
	\eea 
\end{small}
Generalization of this formula for $[A_{n-1}^{(p)}]$ and $[D_{n}^{(p)}]$ minimal models 
has been found long time ago by Fateev and Lukyanov \cite{lukyanov1991additional} by completely different method, 
using Coulomb gas integrals.

To conclude this appendix let us briefly demonstrate how to calculate non-diagonal 
structure constants including the perturbing field $\phi_{-b{\omega_1+\omega_2}}$. 
It follows from the OPE structure that there are two essentially different cases
to investigate: $\alpha_2=\alpha^*_1-b(\omega_1+\omega_2)$ and 
$\alpha_2=\alpha^*_1-2b\omega_1+b\omega_2$ . Other admissible choices are related to 
these two via obvious symmetry of the structure constants under permutation of fields
and the symmetry under particular  Weyl reflection 
$\omega_1\leftrightarrow \omega_2$. Specifying parameters from (\ref{corr})-(\ref{AB_parametrization}) 
and (\ref{sratios}) one can see that the 4-point correlation functions get simplified drastically:
\begin{small}
	\bea
	&&\langle 
	\phi_{\alpha^*_1-b(\omega_1+\omega_2)}(\infty)\phi_{-b \omega_2}(1)\phi_{-b \omega_1}(x)\phi_{\alpha_1}(0)
	\rangle 
	=s_1|x|^{2b(\alpha_1\cdot h_1)}|1-x|^{-\frac{2b^2}{3}}\,,
	\\
	&&\langle 
	\phi_{\alpha^*_1-2b\omega_1+b\omega_2}(\infty)\phi_{-b \omega_2}(1)\phi_{-b \omega_1}(x)\phi_{\alpha_1}(0)
	\rangle 
	=s_2|x|^{2+2b^2+2b(\alpha_1\cdot h_2)}|1-x|^{-\frac{2b^2}{3}}\,.\qquad
	\eea
\end{small}
Comparing s-and t-channels of above correlators with respective OPE`s and 
using known structure constants (\ref{str_simple1})-(\ref{str_simple3}), one obtains 
(\ref{offdiag1}) and (\ref{offdiag2}).

We have seen that investigation of the leading singularities in s-,t- and u-channels
allows to find several structure constants. Similarly, investigating sub-leading singularities 
one can compute also some descendant structure constants. In particular case of our interest 
\bea
\phi_1=\phi^*_0\,,
\qquad
\phi_2=\phi_0\,,
\qquad
\phi_0:=\phi\begin{psmallmatrix}n & n\\n' & n'\end{psmallmatrix}\,.
\eea
the constants  $a$ and $\langle a \bar{a} \rangle$
defined in (\ref{Wphi3}) and (\ref{WWbphi3})   can be evaluated as follows.

The OPE  (\ref{OPEtch}) supplemented also by  the subleading  term becomes
{\footnotesize
\bea 
&\phi_{-b\omega_1}(z)\phi_{-b\omega_2}(1)\sim |z-1|^{-4\Delta(-b\omega_1)}[I]+
|z-1|^{2\Delta(-be_0)-4\Delta(-b\omega_1)}C_{-b\omega_1,-b\omega_2}^{-be_0}\times
\qquad
\\ \nonumber
&\left(1+(z-1)\left(\frac{1}{2}L_{-1}+\beta W_{-1}\right)+...\right)
\left(1+(\bar{z}-1)\left(\frac{1}{2}\bar{L}_{-1}+\beta \bar{W}_{-1}\right)+...\right)
\phi_{-be_0}\,,
\eea
}
where $\beta$ is given in (\ref{beta}).
      Now we can apply this OPE inside the correlation function (\ref{corr}).
      Comparing appropriate powers of $z$ and $\bar{z}$ on both sides we arrive at the expressions  (\ref{a_coef}) and (\ref{aa_coef})
      for  $a$ and $\langle a \bar{a} \rangle$ 
      respectively.

      Similarly for the constants $a_{\alpha \alpha}$ needed for investigation of anomalous $W$-weights for second class  we get the result (\ref{a_alp_alp}).
\section{Hypergeometric functions for small $\epsilon$}
For our purposes we need explicit expressions for the diagonal structure constants (\ref{Cdiag}) in small $\epsilon$ limit. The 
parameters of generalized hypergeometric functions at this limit approach to integer values, which makes their investigation 
rather subtle. In the next subsections we display the results of analysis explicitly.
To illustrate our method  we outline the details of procedure on one specific example 
\begin{small}
	\bea
	\pFq[2]{3}{2}{ 1 - (1 + n') x;, x;,-1 + (2 + n) x}{(n+1)x;,(2-n') x}{1}
	\eea
\end{small}
where 
\[
x=\frac{\epsilon }{3}
\]
is supposed to be small.
Let us examine the expansion (\ref{Def_hypF}) of our hypergeometric function carefully.
The sum of first two  terms behaves as
\begin{small}
	\bea
	\label{first_terms}
	1+\frac{(1-(n'+1) x)((n+2) x-1) }{(n+1) (2-n') x}=
	\frac{1}{(n+1) (n'-2) x}+O\left(1\right)\,.
	\eea
\end{small}
Since $x$ is small, we can choose a large integer $M$  such that 
\bea
\label{M} 
1\ll M\ll \frac{1}{x}\,.
\eea 
Then the terms with $i>M$ can be approximated as
\begin{scriptsize}
	\bea 
	\label{large_M}
	 \frac{(1 - (1 + n') x)_i(x)_i(-1 + (2 + n) x)_i}{((n+1) x)_i  ((2-n') x)_i i!}=
	i^{-x-1} \left(\frac{\Gamma ((n+1) x) \Gamma (-((n'-2) x))}{\Gamma (x) \Gamma ((n+2) x-1) \Gamma (1-(n'+1) x) }+O\left(\frac{1}{i}\right)\right)
	\\ \nonumber
	=
i^{-x-1}	\left(	\frac{n+2}{ (n+1) (n'-2)} +O\left(\frac{1}{i}\right) \right)\,,
	\eea  
\end{scriptsize}
where on second line we have suppressed $O(x)$ correction since, due to  (\ref{M}) it is much smaller than $O(1/i)$. 
Instead, for $i+1$'th term with  $i< M$ we take small $x$ limit
\begin{small}
	\bea
	\frac{(1 - (1 + n') x)_{i+1} (x)_{i+1}(-1 + (2 + n) x)_{i+1}}{((n+1) x)_{i+1} ((2-n') x)_{i+1} (i+1)!}=
	\frac{n+2}{ (n+1) (n'-2)i}+O\left(x\right)\,.
	\eea
\end{small}
Notice that the result is almost the same as (\ref{large_M}) the only difference being that now the factor $i^{-x}$ is missing.
Let us estimate how the sum will be affected if we put this factor by hand. In fact
\bea
\sum_{i=1}^{M-1}
\left(\frac{1}{i}-\frac{1}{i^{1+x}}\right)=
\sum_{i=1}^{M-1}
\left(\frac{x\log i}{i}+O(x^2) \right)<(M-1)x \,,
\eea
which due to our choice of $M$ (\ref{M}) is small.
By definition of Riemann zeta function
\bea
\sum _{i=1}^{\infty } i^{-(x+1)}=\zeta (x+1)=\frac{1}{x}+O(1)\,.
\eea
Thus
\begin{small}
	\bea
	\sum_{i=2}^\infty \frac{(1 - (1 + n') x)_i(x)_i(-1 + (2 + n) x)_i}{((n+1) x)_i  ((2-n') x)_i i!}
	=
	\frac{n+2}{ (n+1) (n'-2)x }+O(1)\,.
	\eea
\end{small}
Adding  contribution of the first two terms (\ref{first_terms}) we find
\begin{small}
	\bea
	\pFq[2]{3}{2}{ 1 - (1 + n') x;, x;,-1 + (2 + n) x}{(n+1)x;,(2-n') x}{1}
	=\frac{n+3}{(n+1) (n'-2)x}+O(1)\,.
	\eea
\end{small}
\subsection{Small $\epsilon$ hypergeometric functions for the second class}
\label{hyp_sec_set}
Here we will be interested in the small $\epsilon$ behavior of the hypergeometric functions needed for 
calculation of diagonal structure constants (\ref{Cdiag}) in case of the second invariant class  (\ref{secund_set_UV}).
For calculation of 
\bea
\nonumber
C^{\begin{psmallmatrix}n & n+1\\n' & n'	\end{psmallmatrix}}
_{\begin{psmallmatrix}1 & 2\\1 &2\end{psmallmatrix},\begin{psmallmatrix}n & n+1\\n' & n'	\end{psmallmatrix}}
\eea
 we need
\begin{scriptsize}
	\bea
	&&\pFq[2]{3}{2} {x,;1-(n'+1) x,;(n+2) x-1}{(n+1) x,;(2-n') x}{1}\approx
	\frac{n+3}{(n+1) (n'-2)x}\,,
	\\
	&&\pFq[2]{3}{2}{x,;x (n+n'),;(n'+1) x}{x (n+n'+3)-1,;n' x+1}{1}\approx
	1\,,
	\\
	&&\pFq[2]{3}{2}{x,;2-x (n+n'+2),;1-n x}{2-(n+1) x,;1-x (n+n'-1)}{1}\approx
	2\,.
	\eea
\end{scriptsize}
For 
\bea
\nonumber
C^{\begin{psmallmatrix}n & n\\n' & n'-1	\end{psmallmatrix}}
_{\begin{psmallmatrix}1 & 2\\1 &2\end{psmallmatrix},\begin{psmallmatrix}n & n\\n' & n'-1	\end{psmallmatrix}}
\eea
we need
\begin{scriptsize}
	\bea
	&&\pFq[2]{3}{2}{x,;-n' x,;(1+n)x}{1+n x,;-1+(3-n') x}{1}\approx
	1\,,
	\\
	&&\pFq[2]{3}{2}{x,;2+ (n+n'-2)x,;1+n' x}{2+(n'-1) x,;1+ (n+n'+1)x}{1}\approx
	2\,,
	\\
	&&\pFq[2]{3}{2}{x,;-(n+n')x,;(1-n) x}{1-n x,;-1-(n+n'-3)x}{1}\approx
	1\,.
	\eea
\end{scriptsize}
For   
\bea
\nonumber
C^{\begin{psmallmatrix}n & n-1\\n' & n'+1	\end{psmallmatrix}}
_{\begin{psmallmatrix}1 & 2\\1 &2\end{psmallmatrix},\begin{psmallmatrix}n & n-1\\n' & n'+1	\end{psmallmatrix}}
\eea
we need
\begin{scriptsize}
	\bea
	&&\pFq[2]{3}{2}{x,;2-(n'+2) x,;1+n x}{2+(n-1) x,;1+(1-n') x}{1}\approx
	2\,,
	\\
	&&\pFq[2]{3}{2}{x,;1+(n+n'-1) x,;(n'+2) x-1}{(n'+1) x,;(n+n'+2) x}{1}\approx
	-\frac{n'+3}{(n'+1) (n+n'+2)x}\,,
	\\
	&&\pFq[2]{3}{2}{x,;1-x (n+n'+1),;(2-n) x-1}{(1-n) x,; (2-n-n')x}{1}\approx
	\frac{n-3}{(n-1) (n+n'-2)x}\,.
	\eea
\end{scriptsize}
\subsection{Small $\epsilon$ hypergeometric functions for the third class}
\label{hyp_third_set}
Here we list the small $\epsilon$ formulae for those hypergeometric functions which enter
in expressions of the structure constants relevant for the third class (\ref{uv10}).\\ 
For the structure constant
\[
C^{\begin{psmallmatrix}n & n+2\\n' & n'-1	\end{psmallmatrix}}
_{\begin{psmallmatrix}1 & 2\\1 &2\end{psmallmatrix},\begin{psmallmatrix}n & n+2\\n' & n'-1	\end{psmallmatrix}}:=C^1_{\,\,1}
\]
 we need the small $\epsilon$ expansions
\begin{scriptsize}
	\bea
	&&\pFq[2]{3}{2}{x,;-n' x,;(n+3) x-2}{(n+2) x-1,;(3-n') x-1}{1}\approx
	\frac{3 n (n'-1)+9 n'-6}{(n+2) (n'-3)}\,,
	\\
	&&\pFq[2]{3}{2}{x,;x (n+n'),;n' x+1}{(n'-1) x+2,;x (n+n'+3)-1}{1}\approx
	\frac{3}{n+n'+3}\,,
	\\
	&&\pFq[2]{3}{2}{x,;2-x (n+n'+2),;2-(n+1) x}{3-(n+2) x,;1- (n+n'-1)x}{1}\approx
	3\,.
	\eea
\end{scriptsize}
For
\[
C^{\begin{psmallmatrix}n & n+1\\n' & n'+1	\end{psmallmatrix}}
_{\begin{psmallmatrix}1 & 2\\1 &2\end{psmallmatrix},\begin{psmallmatrix}n & n+1\\n' & n'+1	\end{psmallmatrix}}:=C^2_{\,\,2}
\]
we need
\begin{scriptsize}
	\bea
	&&\pFq[2]{3}{2}{x,;2-(n'+2) x,;(n+2) x-1}{(n+1) x,;(1-n')x+1}{1}\approx
	-\frac{3}{1+n}\,,
	\\ \nonumber
	&&\pFq[2]{3}{2}{x,;x (n+n'+1)-1,;(n'+2) x-1}{(n'+1) x,;x (n+n'+4)-2}{1}\approx
	\frac{3 \left((n+4) n'+n+n'^2+2\right)}{2 (n'+1) (n+n'+4)}\,,
	\\
	&&\pFq[2]{3}{2}{x,;3-x (n+n'+3),;1-n x}{2-(n+1) x,;2-x (n+n')}{1}\approx
	\frac{3}{2}\,.
	\eea
\end{scriptsize}
For 
\[
C^{\begin{psmallmatrix}n & n+1\\n' & n'-2	\end{psmallmatrix}}
_{\begin{psmallmatrix}1 & 2\\1 &2\end{psmallmatrix},\begin{psmallmatrix}n & n+1\\n' & n'-2	\end{psmallmatrix}}:=C^3_{\,\,3}
\]
we need
\begin{scriptsize}
	\bea
	&&\pFq[2]{3}{2}{x,;(1-n' )x-1,;(n+2) x-1}{(n+1) x,;(4-n') x-2}{1}\approx
	\frac{3}{2} \left(1+\frac{n+2}{(n+1) (n'-4)}\right)\,,
	\\
	&&\pFq[2]{3}{2}{x,;x (n+n'-2)+2,;(n'-1) x+2}{(n'-2) x+3,;x (n+n'+1)+1}{1}\approx
	3\,,
	\\
	&&\pFq[2]{3}{2}{x,;-(n+n')x,;1-n x}{2-(n+1) x,;(3-n-n')x-1}{1} \approx
	-\frac{3}{n+n'-3}\,.
	\eea
\end{scriptsize}
For 
\[
C^{\begin{psmallmatrix}n & n-1\\n' & n'+2	\end{psmallmatrix}}
_{\begin{psmallmatrix}1 & 2\\1 &2\end{psmallmatrix},\begin{psmallmatrix}n & n-1\\n' & n'+2	\end{psmallmatrix}}:=C^8_{\,\,8}
\]
we need
\begin{scriptsize}
	\bea
	&&\pFq[2]{3}{2}{x,;3-(n'+3) x,;n x+1}{(n-1) x+2,;2-n' x}{1}\approx
	\frac{3}{2}\,,
	\\ \nonumber
	&&\pFq[2]{3}{2}{x,;x (n+n'),;(n'+3) x-2}{(n'+2) x-1,;x (n+n'+3)-1}{1}\approx
	\frac{3 (n (n'+3)+n' (n'+4)+2)}{(n'+2) (n+n'+3)}\,,
	\\
	&&\pFq[2]{3}{2}{x,;2-x (n+n'+2),;(2-n) x-1}{(1-n) x,;1- (n+n'-1)x}{1}\approx
	\frac{3}{n-1}\,.
	\eea
\end{scriptsize}
For 
\[
C^{\begin{psmallmatrix}n & n-1\\n' & n'-1	\end{psmallmatrix}}
_{\begin{psmallmatrix}1 & 2\\1 &2\end{psmallmatrix},\begin{psmallmatrix}n & n-1\\n' & n'-1	\end{psmallmatrix}}:=C^9_{\,\,9}
\]
we need
\begin{scriptsize}
	\bea
	&&\pFq[2]{3}{2}{x,;-n' x,;n x+1}{(n-1) x+2,;(3-n') x-1}{1}\approx
	-\frac{3}{n'-3}\,,
	\\
	&&\pFq[2]{3}{2}{x,;x (n+n'-3)+3,;n' x+1}{(n'-1) x+2,;(n +n') x+2}{1}\approx
	\frac{3}{2}\,,
	\\
	&&\pFq[2]{3}{2}{x,;(1-n-n')x-1,;(2-n) x-1}{(1-n )x,;(4-n-n')x-2}{1}\approx
	\frac{3}{2} \left(1+\frac{n-2}{(n-1) (n+n'-4)}\right)\,.
	\nonumber
	\eea
\end{scriptsize}
For 
\[
C^{\begin{psmallmatrix}n & n-2\\n' & n'+1	\end{psmallmatrix}}
_{\begin{psmallmatrix}1 & 2\\1 &2\end{psmallmatrix},\begin{psmallmatrix}n & n-2\\n' & n'+1	\end{psmallmatrix}}:=C^{10}_{\,\,10}
\]
we need
\begin{scriptsize}
	\bea
	&&\pFq[2]{3}{2}{x,;2-(n'+2) x,;(n-1) x+2}{(n-2) x+3,;(1-n') x+1}{1}\approx
	3\,,
	\\
	&&\pFq[2]{3}{2}{x,;x (n+n'-2)+2,;(n'+2) x-1}{(n'+1) x,;(n+n'+1)x+1}{1}\approx
	-\frac{3}{n'+1}\,,
	\\
	&&\pFq[2]{3}{2}{x,;-(n+n')x,;(3-n) x-2}{(2-n) x-1,;(3-n-n')x-1}{1}\approx
	\frac{3 n (n+n'-4)-9 n'+6}{(n-2) (n+n'-3)}\,.
	\eea
\end{scriptsize}
\section{Structure constants for small $\epsilon$}
\label{st_con_3set}
Here we list the structure constants needed for computation of  anomalous dimensions
in the third class.
Let us start with the diagonal ones. From (\ref{Cdiag})   using  the results of \ref{hyp_third_set} for small $\epsilon$ we get
\begin{scriptsize}
	\bea
	\label{C11}
	&& C^1_{\,\,1} \approx \frac{2 (n+3) n'^2+2 n (n+3) n'-n (n+2)}{2 \sqrt{2} (n+1) n' (n+n')}\,,
	\\
	\label{C22}
	&& C^2_{\,\,2} \approx \frac{n^2 (2 n'+1)+2 n (n' (n'+4)+1)+n' (n'+2)}{2 \sqrt{2} n n' (n+n'+1)}\,,
	\\
	\label{C33}
	&& C^3_{\,\,3} \approx \frac{2 n^2 (n'-3)+2 n (n'-3) n'+(n'-2) n'}{2 \sqrt{2} n (n'-1) (n+n')}\,,
	\\
	&& C^4_{\,\,4} \approx C^5_{\,\,5}=C^6_{\,\,6}=\frac{27}{4 \sqrt{2} \left(n^2+n n'+n'^2-3\right)}\,,
	\\
	&& C^7_{\,\,7} \approx \frac{4 \left(n^2+n n'+n'^2-3\right)^3+9 \left(n^2+n n'+n'^2-3\right)^2+3 (n-n')^2 (2 n+n')^2 (n+2 n')^2}{4 \sqrt{2} \left(n^2-1\right) \left(n'^2-1\right) \left((n+n')^2-1\right) \left(n^2+n n'+n'^2-3\right)}\,,
	\\
	\label{C88}
	&& C^8_{\,\,8} \approx \frac{2 n^2 (n'+3)+2 n n' (n'+3)-n' (n'+2)}{2 \sqrt{2} n (n'+1) (n+n')}\,,
	\\
	\label{C99}
	&& C^9_{\,\,9} \approx \frac{n^2 (2 n'-1)+2 n ((n'-4) n'+1)-(n'-2) n'}{2 \sqrt{2} n n' (n+n'-1)}\,,
	\\
	\label{C1010}
	&& C^{10}_{\,\,10} \approx \frac{2 (n-3) n'^2+2 (n-3) n n'+(n-2) n}{2 \sqrt{2} (n-1) n' (n+n')}\,.
	\eea
\end{scriptsize}
The off diagonal structure constants can be obtained from
(\ref{offdiag1}) and (\ref{offdiag2}). Below we list the nonzero structure constants only.
{\footnotesize
\bea
&&C^1_{\,\,2} \approx \frac{1}{2  n'}\sqrt{\frac{(n+2) \left(n'^2-1\right) (n+n'+2)}{2(n+1) (n+n'+1)}}\,,
\\
&&C^1_{\,\,3} \approx \frac{1}{2  (n+n')}\sqrt{\frac{(n+2) (n'-2) \left((n+n')^2-1\right)}{2(n+1) (n'-1)}}\,,
\\
&&C^1_{\,\,4} \approx \frac{3 (n-2)^2 }{4  \left(n^2+n n'+n'^2-3\right)}\sqrt{\frac{(n+2) (n'-1) (n+n'+1)}{2n n' (n+n')}}\,,
\\
&&C^1_{\,\,5}=C^1_{\,\,6} \approx \frac{ (n-2)  (n+2 n')}{4 \left(n^2+n n'+n'^2-3\right)}
\sqrt{\frac{3\left(n^2+n-2\right) (n'+1) (n+n'-1)}{2n (n+1) n' (n+n')}}\,,
\\
&&C^1_{\,\,7}\approx \frac{(n-1) (n'+1) (n+n'-1)  (n+2 n')^2}{4  (n+1) (n'-1) (n+n'+1) \left(n^2+n n'+n'^2-3\right)}
\sqrt{\frac{(n+2) (n'-1) (n+n'+1)}{2n n' (n+n')}}\,.\quad
\eea
}
\begin{scriptsize}
	{\footnotesize

}
\end{scriptsize}
{\footnotesize
\bea
&&C^3_{\,\,4} \approx \frac{3 (n'+2)^2}{4   \left(n^2+n n'+n'^2-3\right)}
\sqrt{\frac{(n'-2) \left(n^2+n n'+n'-1\right)}{2 n n' (n+n')}}\,,
\\
&&C^3_{\,\,5}=C^3_{\,\,6}\approx -\frac{ (n'+2)  (2 n+n')}{4 \left(n^2+n n'+n'^2-3\right)}
\sqrt{\frac{3(n-1) (n'-2) (n'+1) (n+n'+1)}{2 n (n'-1) n' (n+n')}}\,,
\\
&&C^3_{\,\,7} \approx \frac{(n-1) (n'+1) (n+n'+1) (2 n+n')^2}{4  (n+1) (n'-1) (n+n'-1)  \left(n^2+n n'+n'^2-3\right)}
\sqrt{\frac{(n'-2) \left(n^2+n n'+n'-1\right)}{2 n n' (n+n')}}\,,
\\
&&C^3_{\,\,9}\approx \frac{1}{2  n}\sqrt{\frac{\left(n^2-1\right) (n'-2) (n+n'-2)}{2 (n'-1) (n+n'-1)}}\,,
\eea
}
{\footnotesize
\bea
&&C^4_{\,\,7}\approx \frac{27}{4 \sqrt{2} \left(n^2+n n'+n'^2-3\right)}\,,
\\
&&C^4_{\,\,8} \approx \frac{3 (n'-2)^2 }{4  \left(n^2+n n'+n'^2-3\right)}\sqrt{\frac{(n-1) (n'+2) (n+n'+1)}{2 n n' (n+n')}}\,,
\\
&&C^4_{\,\,9} \approx \frac{3  (n+n'+2)^2}{4  \left(n^2+n n'+n'^2-3\right)}\sqrt{\frac{(n-1) (n'-1) (n+n'-2)}{2 n n' (n+n')}}\,,
\\
&&C^{4}_{\,\,10} \approx \frac{3 (n+2)^2 }{4 \left(n^2+n n'+n'^2-3\right)}\sqrt{\frac{(n-2) (n'+1) (n+n'-1)}{2 n n' (n+n')}}\,,
\eea
}
{\footnotesize
\bea
&&C^5_{\,\,6}\approx \frac{27}{4 \sqrt{2} \left(n^2+n n'+n'^2-3\right)}\,,
\\
&&C^5_{\,\,7} \approx \frac{3  (n'-n) (2 n+n') (n+2 n')}{2 \left(n^2+n n'+n'^2-3\right)}
\sqrt{\frac{3}{2(n^2-1) (n'^2-1)  ((n+n')^2-1) }}\,,
\\
&&C^5_{\,\,8} \approx -\frac{ (n'-2) (2 n+n') }{4 \left(n^2+n n'+n'^2-3\right)}
\sqrt{\frac{3(n+1) \left(n'^2+n'-2\right) (n+n'-1)}{2 n n' (n'+1) (n+n')}}\,,
\\
&&C^5_{\,\,9} \approx \frac{ (n'-n)  (n+n'+2)}{4 \left(n^2+n n'+n'^2-3\right)}
\sqrt{\frac{3(n+1) (n'+1) (n+n'-2) (n+n'+1)}{2 n n' (n+n'-1) (n+n')}}\,,
\\
&&C^5_{\,\,10} \approx \frac{ (n+2)  (n+2 n')}{4 \left(n^2+n n'+n'^2-3\right)}
\sqrt{\frac{3(n-2) (n+1) (n'-1) (n+n'+1)}{2 (n-1) n n' (n+n')}}\,,
\eea
}
{\footnotesize
\bea
&&C^6_{\,\,7}\approx \frac{3 (n'-n) (2 n+n') (n+2 n')}{2  \left(n^2+n n'+n'^2-3\right)}
\sqrt{\frac{3}{2(n^2-1) (n'^2-1)  ((n+n')^2-1)}} \,,
\\
&&C^6_{\,\,8}\approx -\frac{ (n'-2) (2 n+n') }{4 \left(n^2+n n'+n'^2-3\right)}
\sqrt{\frac{3(n+1) \left(n'^2+n'-2\right) (n+n'-1)}{2 n n' (n'+1) (n+n')}}\,,
\\
&&C^6_{\,\,9}\approx \frac{ (n'-n) (n+n'+2)}{4 \left(n^2+n n'+n'^2-3\right)}
\sqrt{\frac{3(n+1) (n'+1) (n+n'-2) (n+n'+1)}{2 n n' (n+n'-1) (n+n')}} \,,
\\
&&C^6_{\,\,10}\approx \frac{ (n+2)  (n+2 n')}{4 \left(n^2+n n'+n'^2-3\right)}
\sqrt{\frac{3 (n-2) (n+1) (n'-1) (n+n'+1)}{2 (n-1) n n' (n+n')}}\,,
\eea
}
{\footnotesize
\bea
C^7_{\,\,8}\approx \frac{(n+1) (n'-1) (n+n'-1)  (2 n+n')^2}{4  (n-1) (n'+1) (n+n'+1) \left(n^2+n n'+n'^2-3\right)}
\sqrt{\frac{(n-1) (n'+2) (n+n'+1)}{2 n n' (n+n')}}\,,
\\
C^7_{\,\,9}\approx \frac{(n+1) (n'+1) (n-n')^2  (n+n'+1)}{4  (n-1) (n'-1) (n+n'-1) \left(n^2+n n'+n'^2-3\right)}
\sqrt{\frac{(n-1) (n'-1) (n+n'-2)}{2 n n' (n+n')}}\,,
\\
C^7_{\,\,10}\approx \frac{(n+1) (n'-1)  (n+n'+1) (n+2 n')^2}{4  (n-1) (n'+1) (n+n'-1) \left(n^2+n n'+n'^2-3\right)}
\sqrt{\frac{(n-2) (n'+1) (n+n'-1)}{2 n n' (n+n')}}\,,
\eea
}
{\footnotesize
\bea
C^8_{\,\,10}\approx \frac{1}{2 \sqrt{2} (n+n')}\sqrt{\frac{(n-2) (n'+2) \left((n+n')^2-1\right)}{(n-1) (n'+1)}}\,,
\eea
}
{\footnotesize
\bea
C^9_{\,\,10}\approx 
\frac{1}{2 \sqrt{2} n'}\sqrt{\frac{(n-2) \left(n'^2-1\right) (n+n'-2)}{(n-1) (n+n'-1)}}\,.
\eea
}
\section{Representation of $\phi_i^{IR}\phi_j$  in terms of direct product WZNW models in case of second class}
\label{app_dom_wall}
We have already shown in the main text (see eq. (\ref{psi21})) that
\bea
\nonumber
|11 \rangle= |\lambda_{n-1,n'}\rangle|\lambda_{2,1}\rangle |\lambda_{2,1}\rangle\,.
\eea 
Next  consider  $\phi^{IR}\begin{psmallmatrix}n-1 & n\\n' & n'	\end{psmallmatrix}\phi^{UV}\begin{psmallmatrix}n & n\\n' & n'-1	\end{psmallmatrix}$:
\bea
\Delta^{IR}\begin{psmallmatrix}n-1 & n\\n' & n'	\end{psmallmatrix}=
h_{k-1}(\lambda_{n-1,n'})+h_1(\lambda_{2,1})-h_k(\lambda_{n,n'}),
\\
\Delta^{UV}\begin{psmallmatrix}n & n\\n' & n'-1	\end{psmallmatrix}=
h_{k}(\lambda_{n,n'})+h_1(\lambda_{2,1})-h_{k+1}(\lambda_{n,n'-1}).
\eea
It os easy to check that
\bea
\lambda_{n-1,n'}+\lambda_{2,1}+\lambda_{2,1}+(-1,0,1)^T=\lambda_{n,n'-1}\,,
\\
\lambda_{n-1,n'}+\lambda_{2,1}=\lambda_{n,n'}\,.
\eea
The second equality implies that one should consider only such states
which belong to the irreducible representation of the combined current $K=E+J$ with highest weight 
$\lambda_{n-1,n'}+\lambda_{2,1}$. Taking into account also the constraint coming from the first 
equality one is lead to the ansatz
\begin{small}
	\bea
	|12 \rangle=
	\left(
	a_1K_{31}+a_2\tilde{J}_{31}+a_3K_{32}K_{21}+a_4K_{32}\tilde{J}_{21}
	\right)
	|\lambda_{n-1,n'}\rangle|\lambda_{2,1}\rangle  |\lambda_{2,1}\rangle\,.
	\eea
\end{small}
Now we can impose the condition that $|12 \rangle$ should be a  highest weight state 
of total current
\bea
\label{hwsCon}
\left (K_{ij}+\tilde{J}_{ij}\right)|12 \rangle=0
\quad {\rm with} \quad i<j\,.
\eea
This fixes the coefficients $a$ up to an overall constant which we will choose such that  $a_1=1$.
Here is the result
{\footnotesize
	\bea
	\label{psi22}
	&|12 \rangle =
	\big(
	(E+J)_{31}-\frac{(1-n-n')(1-n')}{n'}\tilde{J}_{31}- \frac{1}{n'}(E+J)_{32}(E+J)_{21}-
	\\
	\nonumber
	& \qquad \qquad \qquad \qquad \qquad \qquad
	-\frac{ 1-n-n'}{n'}(E+J)_{32}\tilde{J}_{21}	\big)
	|\lambda_{n-1,n'}\rangle|\lambda_{2,1}\rangle |\lambda_{2,1}\rangle .
	\eea
}
The other states are obtained in a similar fashion. The results are listed below.\\
\begin{small}
	\bea
	\label{psi23}
	&|13 \rangle= \left((E+J)_{21}+(1-n)\tilde{J}_{21}\right) 
	|\lambda_{n-1,n'}\rangle|\lambda_{2,1}\rangle |\lambda_{2,1}\rangle\,,
	\eea
\end{small} 
{\footnotesize
	\bea
	\label{psi_11}
	&	|21 \rangle=
	\left(
	E_{31}-\frac{n'(n+n')}{1+n'}J_{31}-\frac{1}{1+n'}E_{32}E_{21}+\frac{n+n'}{n'+1}E_{32}J_{21}
	\right)
	|\lambda_{n,n'+1}\rangle |\lambda_{2,1}\rangle \,,
	\qquad\qquad\qquad\quad\,
	\\
	\label{psi12}
	&|22\rangle =
	\left(
	(E+J)_{31}-\frac{(1-n-n')(1-n')}{n'}\tilde{J}_{31}- \frac{1}{n'}(E+J)_{32}(E+J)_{21}-
	\frac{ 1-n-n'}{n'}(E+J)_{32}\tilde{J}_{21}
	\right)
	\\
	&\qquad\qquad\qquad\times\left(
	E_{31}-\frac{n'(n+n')}{1+n'}J_{31}-\frac{1}{1+n'}E_{32}E_{21}+\frac{n+n'}{n'+1}E_{32}J_{21}
	\right)
	|\lambda_{n,n'+1}\rangle|\lambda_{2,1}\rangle|\lambda_{2,1}\rangle\,,
	\nonumber
	\\
	\label{psi13}
	&	|23 \rangle=
	\left((E+J)_{21}+(1-n)\tilde{J}_{21}\right)\times 
	\qquad \qquad \qquad \qquad\qquad \qquad \qquad\qquad \qquad\qquad\qquad \quad
	\\ \nonumber
	&\qquad\qquad	\times
	\left(
	E_{31}-\frac{n'(n+n')}{1+n'}J_{31}-\frac{1}{1+n'}E_{32}E_{21}+\frac{n+n'}{n'+1}E_{32}J_{21}
	\right)
	|\lambda_{n,n'+1}\rangle|\lambda_{2,1}\rangle|\lambda_{2,1}\rangle\,,
	\eea	
}
{\footnotesize
	\bea
	\label{psi31}
	&|31 \rangle=(E_{21}-nJ_{21})
	|\lambda_{n+1,n'-1}\rangle|\lambda_{2,1}\rangle |\lambda_{2,1}\rangle\,,
	\qquad\qquad\qquad\qquad\qquad\qquad\qquad\qquad\qquad\qquad\,
	\\
	\label{psi32}
	&|32 \rangle =
	\left(
	(E+J)_{31}-\frac{(1-n-n')(1-n')}{n'}\tilde{J}_{31}- \frac{1}{n'}(E+J)_{32}(E+J)_{21}-
	\frac{ 1-n-n'}{n'}(E+J)_{32}\tilde{J}_{21}
	\right)
	\nonumber   \\ 
	&\qquad\qquad\qquad\qquad\qquad\qquad \qquad\qquad \qquad\qquad\times
	(E_{21}-n J_{21})
	|\lambda_{n+1,n'-1}\rangle|\lambda_{2,1}\rangle |\lambda_{2,1}\rangle\,,
	\\
	\label{psi33}
	&|33 \rangle=\left((E+J)_{21}+(1-n)\tilde{J}_{21}\right)
	(E_{21}-n J_{21})
	|\lambda_{n+1,n'-1}\rangle|\lambda_{2,1}\rangle |\lambda_{2,1}\rangle\,.
	\qquad\qquad\qquad\qquad\quad
	\eea
}
Having above explicit expressions we have computed ratio of scalar  products $\langle \widetilde{ij}	|ij \rangle$ and
$\langle ij	|ij \rangle$.

From  (\ref{psi21}) we simply get
\bea
\frac{\langle \widetilde{11}	|11 \rangle}{\langle 11	|11 \rangle}=1\,.
\eea
From (\ref{psi22}) 
{\footnotesize
	\bea
	\label{psitilde22}
	&\langle \widetilde{12} | =
	\left(
	(E+\tilde{J})_{13}-\frac{(1-n-n')(1-n')}{n'}J_{13}- \frac{1}{n'}(E+\tilde{J})_{12}(E+\tilde{J})_{23}-
	\frac{ 1-n-n'}{n'}(E+\tilde{J})_{23}J_{12}
	\right)
	\nonumber   \\ 
	&
	|\lambda_{n-1,n'}\rangle|\lambda_{2,1}\rangle |\widetilde{\lambda_{2,1}}\rangle .
	\eea
}
so
\bea
\frac{\langle \widetilde{12}	|12 \rangle}{\langle 12	|12 \rangle}=
\frac{1}{1-n-n'}\,.
\eea
Let us list the other cases
\begin{itemize}
\item  From (\ref{psi23}) we get
\bea
\frac{\langle \widetilde{13}	|13 \rangle}{\langle 13	|13 \rangle}=
\frac{1}{1-n}\,.
\eea
\item  From  (\ref{psi_11}) we get  
\bea
\frac{\langle \widetilde{21}	|21 \rangle}{\langle 21	|21 \rangle}=
\frac{1}{n+n'+1}\,.
\eea
\item  From (\ref{psi12})  we get
\bea
\label{psi12psi12}
\frac{\langle \widetilde{22}	|22 \rangle}{\langle 22	|22 \rangle}=1\,.
\eea
\item  From  (\ref{psi13}) we get
\bea
\frac{\langle \widetilde{23}	|23 \rangle}{\langle 23	|23 \rangle}=
\frac{1}{n'+1}\,.
\eea

\item  From (\ref{psi31}) we get
\bea
\frac{\langle \widetilde{31}	|31 \rangle}{\langle 31	|31 \rangle}=
\frac{1}{n+1}\,.
\eea
\item  From (\ref{psi32}) we got
\bea
\frac{\langle \widetilde{32}	|32 \rangle}{\langle 32	|32 \rangle}=
\frac{1}{1-n'}\,.
\eea
\item  From (\ref{psi33}) we get
\bea
\frac{\langle \widetilde{33}	|33 \rangle}{\langle 33	|33 \rangle}=1\,.
\eea
\end{itemize}
\end{appendix}
\bibliographystyle{JHEP}
\providecommand{\href}[2]{#2}
\providecommand{\href}[2]{#2}\begingroup\raggedright\endgroup

\end{document}